\documentclass{jfm}
\usepackage[normalem]{ulem} 
\usepackage{color}
\usepackage{graphicx}
\usepackage{natbib}
\usepackage{psfrag}

\usepackage{subfigure}
\usepackage{amsmath}

\title{Interaction of two high Reynolds number axisymmetric turbulent
  wakes}

\author[S. Klein, M. Obligado and J.C. Vassilicos]%
       {M. Obligado\aff{1}, S. Klein\aff{2} and J.C. Vassilicos\aff{3}
         \corresp{Email address for correspondence:  martin.obligado@univ-grenoble-alpes.fr \& john-christos.vassilicos@centralelille.fr}}

\affiliation{\aff{1} Universit\'{e} Grenoble Alpes, CNRS, Grenoble-INP, LEGI, F-38000, Grenoble, France 
\aff{2} Institute of Fluid Mechanics, TU Braunschweig, Braunschweig, Germany
\aff{3} Univ. Lille, CNRS, ONERA, Arts et M\'etiers ParisTech, Centrale Lille, UMR 9014 - LMFL - Laboratoire de M\'ecanique des fluides de Lille - Kamp\'e de Feriet, F-59000 Lille, France}

\pubyear{2010}
\volume{650}
\pagerange{119--126}
\date{?; revised ?; accepted ?. - To be entered by editorial office}

\begin{document}
\sloppy 

\maketitle
\begin{abstract}

The interaction between turbulent axisymmetric wakes plays an
important role in many industrial applications, notably in the
modelling of wind farms. While the non-equilibrium high Reynolds
number scalings present in the wake of axisymmetric plates has been
shown to modify the averaged streamwise scalings of individual wakes,
little attention has been paid to their consequences in terms of wake
interactions. We propose an experimental setup that tests the presence
of non-equilibrium turbulence using the streamwise variation of
velocity fluctuations between two bluff bodies facing a laminar
flow. We have studied two different sets of plates (one with regular
and another with irregular peripheries) with hot-wire anemometry in a
wind tunnel. By acquiring streamwise profiles for different plate
separations and identifying the wake interaction length for each
separation it is possible to show that the interaction between them is
consistent with non-equilibrium scalings. This work also generalises
previous studies concerned with the interaction of plane wakes to
include axisymmetric wakes. We find that a simple mathematical
expression for the wake interaction length based on non-equilibrium
turbulence scalings can be used to collapse the streamwise
developments of the second, third and fourth moments of the streamwise
fluctuating velocity.

\end{abstract}

\begin{keywords}
Wakes, turbulence theory, experimental fluid mechanics
\end{keywords}

\section{Introduction} \label{intro}

Recently, flow regions with non-equilibrium high Reynolds number
turbulence at odds with usual Richardson-Kolmogorov phenomenology have
been discovered in a number of turbulent
flows~(\cite{vassilicos15,chongsiripinyo2020decay}), in particular
axisymmetric and self-preserving turbulent wakes of plates with and
without irregular edges. These regions are characterised by streamwise
evolutions of the mean flow profiles which have only recently been
documented and partially understood in experiments~(\cite{Nedic2013,
  obligado2015}). The presence of a different set of scalings has many
consequences, such as variations in the turbulent entrainment in
free-shear
flows~(\cite{zhou2017related,van2021unified,cafiero2019non,cafiero2020non})
and on eddy viscosity models~(\cite{cafiero2020length}), among
others. Furthermore, these regions can extend as far as about 100
plate characteristic lengths (defined as $\sqrt{{\cal{A}}}$, with
${\cal{A}}$ the frontal area of the plate) in the streamwise
direction. A further study by ~\cite{dairay15}, critically revised the
classical theory of high Reynolds axisymmetric turbulent
wakes~(\cite{townsend80,george89}) to encompass these new
scalings. Both direct numerical simulations and experiments were found
to agree with the theory.

In this work we focus on the interaction of turbulent axisymmetric
wakes generated by two bluff bodies. This is an important
configuration, present for instance in arrays of wind or marine tidal
turbines, and the interaction of the two wakes can be expected to
involve non-equilibrium turbulence. While some experiments in wind
tunnel controlled conditions have been performed
recently~(\cite{neunaber2020distinct,neunaber2021investigation,
  scott2020wind}), no attention has been paid to the relation between
the energy cascade of the turbulent flow and the wake interaction
length (defined as the streamwise distance at which two wakes merge).

The non-equilibrium predictions and the classical predictions rely on
axisymmetry of turbulence wake statistics, self-preservation of
$(U_{\infty}-U)/u_0$ (with $U_\infty$ the freestream velocity, $U$ the
streamwise mean velocity and $u_0=U_\infty-U_0$ the centreline
velocity deficit), turbulent kinetic energy $K$,
turbulence dissipation $\varepsilon$ and the sum of production and
turbulent transport, and on a scaling law for the centreline
turbulence dissipation (\cite{dairay15,cafiero2019non}). Both sets of
  predictions are obtained from the Reynolds averaged streamwise
  momentum and turbulent kinetic energy equations leading to a closed
  set of equations for $u_0 (x)$ and the wake half width
  $\delta(x)$. The equilibrium predictions for axisymmetric turbulent
  wakes \citep[see][]{townsend80,george89} for the streamwise
  evolution (along $x$) of $u_0$ and $\delta$ are
\begin{equation} \label{eq3}
u_{0} (x) = A U_{\infty} \left((x-x_{0})/\theta\right)^{-2/3},
\end{equation}
\begin{equation} \label{eq4}
\delta (x) = B \theta \left((x-x_{0})/\theta\right)^{1/3},
\end{equation}
where $A$ and $B$ are dimensionless constants, $\theta$ the momentum
thickness and $x_0$ a virtual origin. The momentum thickness $\theta$
is defined by $\theta^2= \frac{1}{U_{\infty}^{2}} \int_0^\infty
U_\infty \left(U_\infty-U \right) r dr$ which is constant with $x$,
and the wake's width is here characterised by the integral wake's half
width defined by $\delta^2(x)= \frac{1}{u_0}\int_0^\infty
\left(U_\infty-U \right) r dr$.

On the other hand, the non-equilibrium predictions are
\begin{equation} \label{eq1}
u_{0} (x) = A U_{\infty}
\left((x-x_{0})/\theta\right)^{-1},
\end{equation}
\begin{equation} \label{eq2}
\delta (x) = B \theta \left((x-x_{0})/\theta\right)^{1/2}.
\end{equation} 

The only difference between equilibrium and non equilibrium scalings
is in the scaling of the centreline value of
$\varepsilon$~(\cite{vassilicos15}), that will be different according
to the nature of the energy cascade. It is then possible to model the
interaction between wakes via the streamwise scaling of $\delta$. It
is expected that within the equilibrium cascade, the wake interaction
length $x^*$ defined by $2\delta (x^{*}) \propto S$ (see
\cite{Mazellier2010,gomes2012}) will evolve as $S^3$, with $S$ the
separation between the centre of the plates. Accordingly, the presence
of non-equilibrium energy cascade implies that $x^* \propto S^2$.

We present experimental evidence that the non-equilibrium theory
in~\cite{dairay15} properly models the interaction of two axisymmetric
wakes, for plates with both regular and irregular edges. We show that,
by having knowledge of the values of the wake width $\delta$ and the
centreline velocity deficit $u_0$ for a single wake, it is possible to
predict the $x^*$ which quantifies the position where the wakes
meet. Furthermore, it is also possible to predict the intensity of the
fluctuations at that particular point.  For this purpose, we propose
an experimental setup where streamwise profiles of streamwise
fluctuating velocities are acquired via hot-wire anemometry in a
wind tunnel. We have tested two different sets of plates, one with
square regular and another with irregular edges.

This work generalizes previous studies on the interaction of plane
wakes (see~\cite{gomes2012}) to include axisymmetric wakes. We find
that the derivation of the wake interaction length proposed in this
cited work
can be used to collapse the streamwise development of the first three
fluctuating velocity moments. Our results suggests that
non-equilibrium scalings
are in good agreement with the interaction of wakes for both sets of
plates studied.

\section{Experimental setup}

\begin{figure}
\centering
\subfigure[\label{abb_01_expsetup}]{\includegraphics[width=0.55\textwidth]{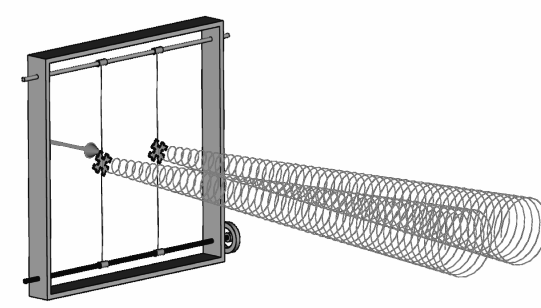}}
\subfigure[\label{abb_02_fractal}]{\includegraphics[width=0.35\textwidth]{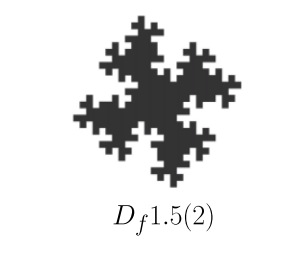}}
\caption{(a) Sketch of the experimental wind tunnel set-up; (b) Irregular plate used herein. They are the second iteration of a fractal plate with dimension $D_f=1.5$ and a square initial pattern, as characterised by~\cite{Nedic2013b}.} \label{fig:fig1}
\end{figure}

The experiments were conducted in a low turbulence wind tunnel with a
test-section of $3\times3$~ft\textsuperscript{2} ($0.91 \times
0.91$~m\textsuperscript{2}) and $4.25$~m long. The bluff plates were
placed at the beginning of the test-section. Figure \ref{fig:fig1}a
presents a sketch of the wind tunnel set-up. Two solid iron bars (with
a diameter of $16$~mm) fixed to the wind tunnel sidewalls close to
ceiling and bottom served as main support for the plates. The plates,
each fixed to a thin iron rod (diameter $1.5$~mm, length $750$~mm),
were connected to the main support by two pairs of T-like
casings. These casings were attached movably to the main support bars
to allow changes in plate separation. By means of small grub screws,
the T-casings could be fixed to the main support bars.

We tested two different sets of plates: two with regular square
peripheries and other two with irregular peripheries. These irregular
plates, shown in figure \ref{fig:fig1}b, are identical to those used
in some previous works~(\cite{dairay15, obligado2015,
  Nedic2013}). They have a reference length of
$L_b=\sqrt{\cal{A}}=64$~mm (with $\cal{A}$ the frontal area of the
plates) and a thickness of 1.25~mm. Only interactions of the turbulent
wakes of the same type of plate were studied, therefore we did
measurements with two regular and two irregular plates only. The
plates were located vertically in the symmetry plane of the wind
tunnel, normal to the laminar freestream velocity. In spanwise
direction, they were equally spaced to the streamwise symmetry
plane. The total blockage of the set-up remains low, close to 4.3\%,
and therefore we do not consider any blockage corrections to our
results.

Eleven different plate separations $S$ were tested: 230, 240, 250, 260, 270,
280, 285, 290, 295, 300 and 305~mm. A right-handed coordinate system
serves as reference, with $+x$ pointing downstream, $+y$ pointing to
the bottom, and $+z$ pointing in spanwise direction. The origin is set
at the centre point of the wind tunnel at the streamwise position of
the plates. Thus, $x$ marks the streamwise distance to the plane of
the plates. Freestream velocity was kept constant at $U_\infty=10$~m/s
throughout the tests and was controlled and stabilised with a PID
feedback system using the static pressure difference across the $9:1$
wind tunnel contraction and the temperature inside the test-section
measured half way along it. At that velocity, the fluctuations around
the mean are below 0.1\% when the test-section is empty.

Hot-wire anemometry measurements were conducted downstream of the pair
of plates using a Dantec Dynamics 55P01 single hot wire, driven by a Dantec StreamLine CTA
system. The probe has a Pt-W wire, 5$\,\mu$m in diameter, 3 mm long
with a sensing length of 1.25 mm. The probe was placed with a levelling laser at the wind
tunnel centre line and velocity profiles in streamwise direction were
recorded with intervals of 10~mm. The probe can be located as close as
100~mm from the plates and up to 3020~mm away from them. For each
probe location, the acquisition time was 60~s with a sampling
frequency of 20~kHz. The traverse system is modular, allowing to
automatize the acquisition of streamwise profiles over a 540~mm
span. It can then be moved to cover different regions of the test
section. Therefore, to measure sufficiently long streamwise distances,
two or more profiles were recorded for each plate separation $S$. For
the particular $S=285$~mm case, a set of streamwise profiles covering
the entire test section were performed for both kinds of
plate. Therefore, for $S=285$~mm, we have access to the whole
streamwise evolution of the velocity temporal signal of the streamwise velocities.
 
To ensure continuity between individual profiles, they overlapped for
approximately 100~mm. At the beginning of each single profile, a
vertical wake profile was acquired between -250 mm$\,< y <\,$250 mm
with $\Delta y=20$ mm to verify $U_\infty$ (acquisition time 30 s,
frequency 20 kHz), and a new calibration of the hot-wire was made to
account for possible thermal drifts of the wind tunnel. Calibrations
were made with a reference from a pitot tube located 50 mm below the
hot-wire and for 9 equispaced velocities between 5 and
15~m/s. Temperature was monitored during the streamwise profiles, so
it never changed by more than 0.2$^\circ$C.

The acquisition time of the centreline single wire measurements being
$60$ sec, an order of 1,000 integral time scales at each streamwise
position were recorded thereby allowing good large-scale resolution.
The Kolmogorov frequency was always smaller than half our sampling
frequency (which is 20kHz) and the fluctuations are always below
10$\%$ (see figure \ref{flow}a). 

\section{Results -  Plates with irregular edges}

\subsection{Streamwise profiles and wake scale $x_{12}$}

\begin{figure}
\centering
\includegraphics[width=0.7\textwidth]{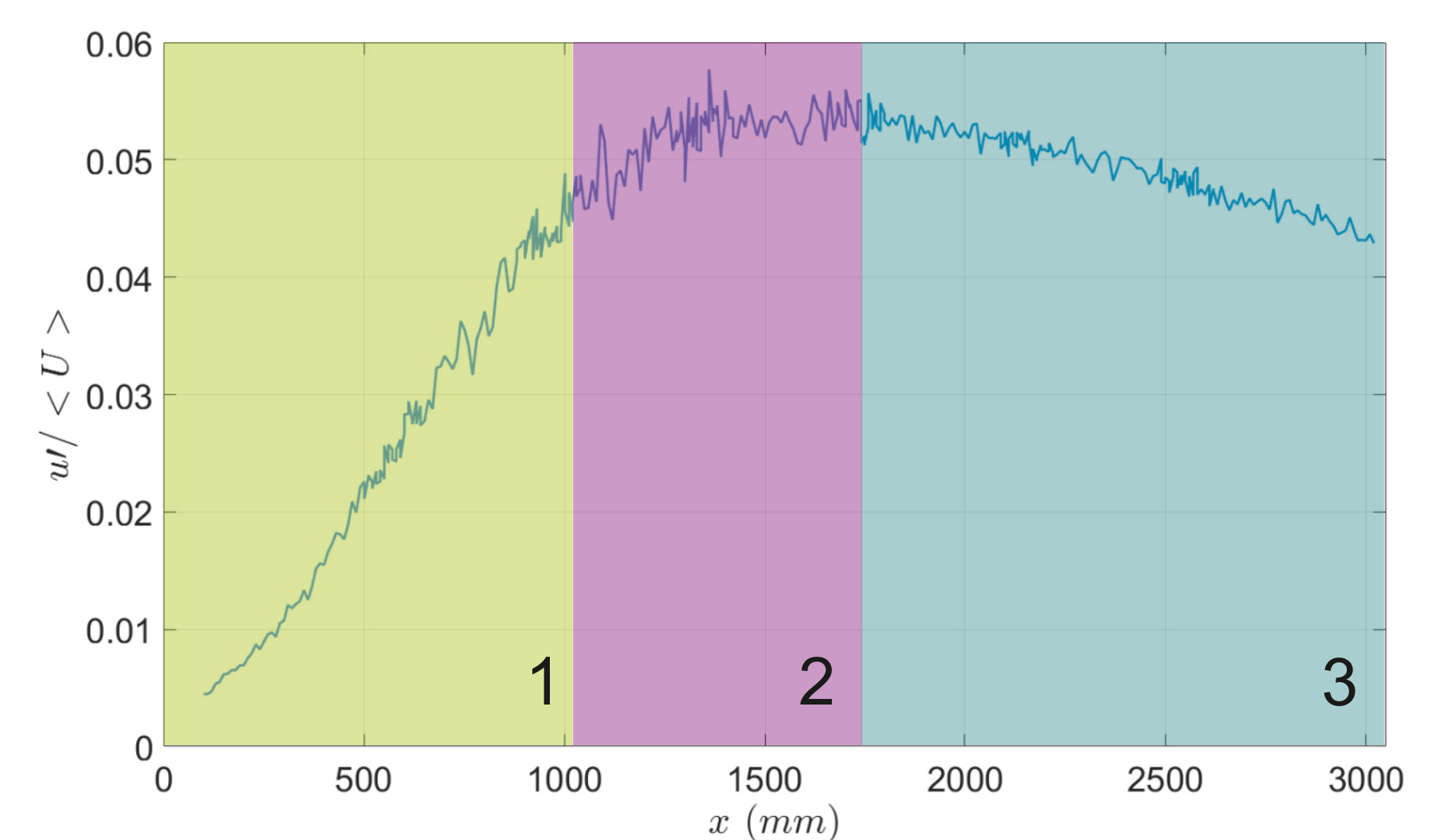}
\caption{Streamwise distribution of  $u\prime/\langle U(x)\rangle$ for the irregular plates at $S=285mm$.}
\label{abb_fractal_U_Ustd}
\end{figure}

For a plate separation of 285 mm, streamwise velocity measurements
along the entire test-section have been conducted. To capture this
length, eleven single velocity profiles have been recorded. These
measurements yield directly the local streamwise mean velocity $\langle
U(x)\rangle$ and the corresponding fluctuating velocity around this mean
$u(t)$. Figure \ref{abb_fractal_U_Ustd} shows the streamwise
distribution of the velocity fluctuations $u\prime/\langle
U(x)\rangle$, with $u\prime$ the standard deviation of $u(t)$.

From figure~\ref{abb_fractal_U_Ustd} we identify three different
regimes in the flow (highlighted in the figure with different colours
and labelled 1, 2 and 3). In region 1, both wakes have not met yet. It
is important to remark that the boundaries of the wake are of
statistical nature, and the flow can be affected by the generator at
radial distances much higher than $\delta$. This is the reason why the
fluctuations show a monotonic increase in this region. Then, the
profile of $u\prime/U_\infty(x)$ suggests that in region 2 both wakes
start to significantly interact and their boundaries meet more often
than not, so that the velocity fluctuations reach a peak. This change
of regime is characterised by a sudden loss of linearity (present for
example in the range $x\subset[500;900]$~mm), as better detailed
below. Further downstream, in region 3, both wakes are fully merged
and the fluctuations exhibit a monotonic decrease.

\begin{figure}
\centering
\includegraphics[width=0.95\textwidth]{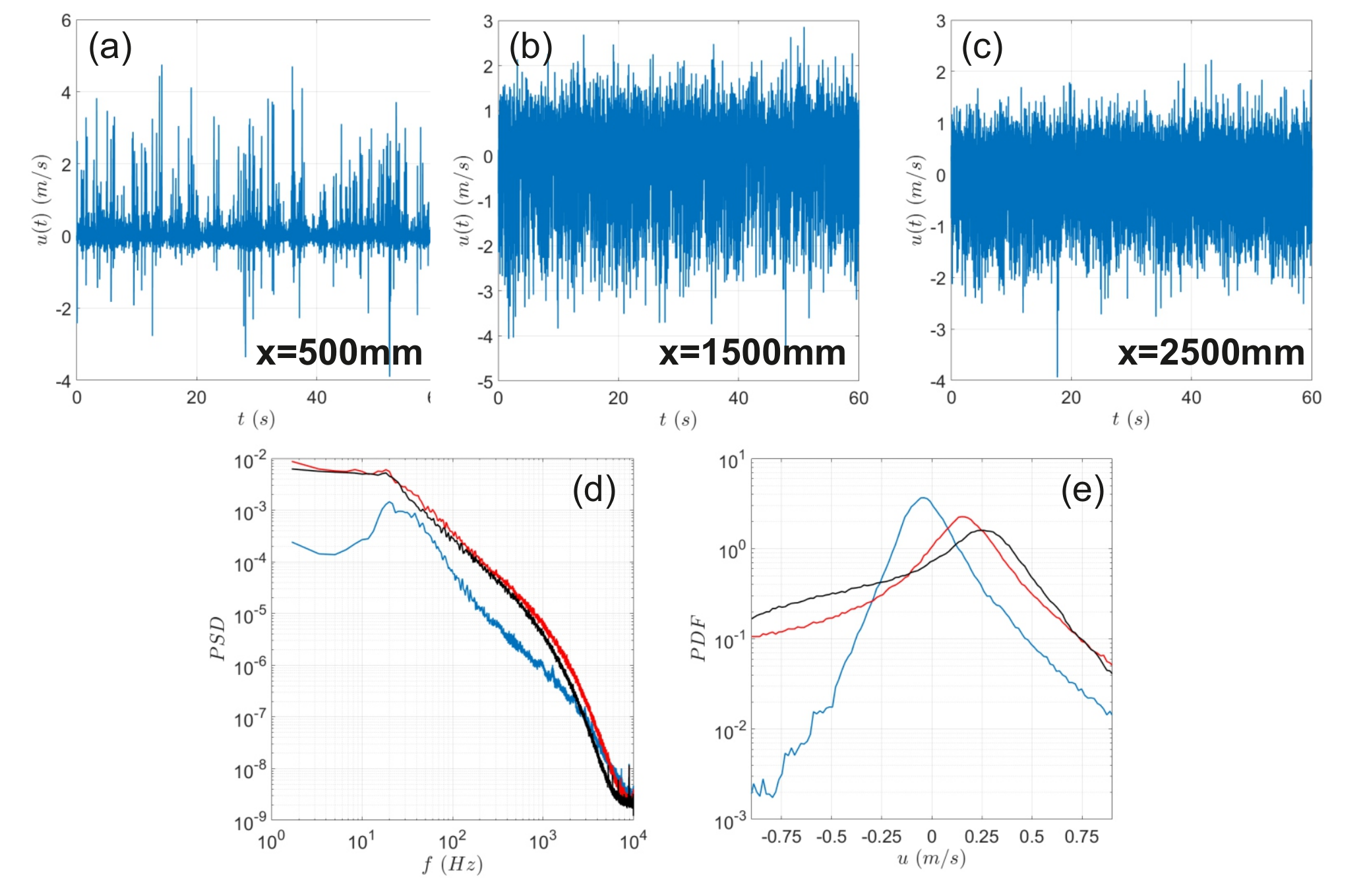}
\caption{Streamwise fluctuating velocity time signal $u(t)$ and its spectra and PDF at different
  locations for irregular plates at $S=285$~mm. Time signal at
  $x_1=500$~mm (a), $x_2=1500$~mm (b) and $x_3=2500$~mm (c). Spectra
  at the three different locations (d), blue lines corresponds to
  $x_1$, the red one to $x_2$ and the black one to $x_3$. PDF of
  $u(t)$ (e) at the same locations and represented with the same
  colours as in (d). At $x_1$, the signal has a skewness of 3.0 and a
  flatness of 39.0. At $x_2$ they are -1.9 and 8.5,
  respectively. Finally, the signal at $x_3$ has a skewness of -1.1
  and a flatness of 4.4.}
\label{flow}
\end{figure}

For a better comprehension of the interactions regimes, we study the
properties of the flow at three representative streamwise locations:
$x_1=500$~mm, $x_2=1500$~mm and $x_3=2500$~mm. Figures
\ref{flow}$a,b~\&c$ show the time-signal of the streamwise fluctuating velocity $u (t)$ at these
points. At $x_1$, the velocity signal is almost flat, with some
extreme events. The spectrum is not turbulent yet (figure
\ref{flow}d), while the probability density function (PDF) of $u(t)$
(figure \ref{flow}e) is far from Gaussian, and positively
skewened. This behaviour is reminiscent of the findings
in~\cite{Laizet2015} for a turbulent grid flow, very close to the grid
bars. Regions 2 and 3 show a better developed turbulent spectrum and a
PDF which is still non-Gaussian, but has become negatively skewed.
Figures \ref{abb_fractal_U_Ustd} and \ref{flow} therefore suggest that
the point $x_{12}$ which demarcates between regions 1 and 2 might be
related with the interaction of the two wakes and might be a good
candidate for the wake interaction length $x^*$.
 
\begin{figure}
\centering
\includegraphics[width=0.6\textwidth]{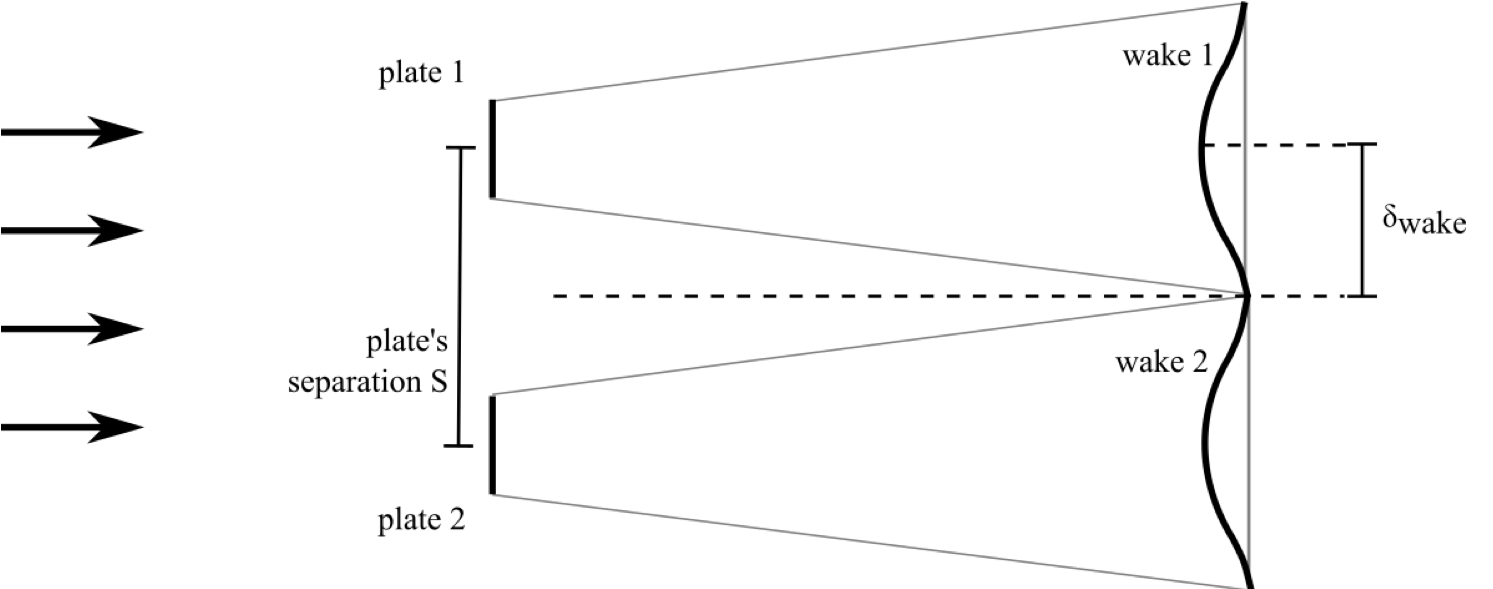}
\caption{Sketch of the interaction of both wakes when $0.5\cdot S = n\cdot\delta$.}
\label{abb:sketch_wake_inter}
\end{figure}

In the following, we test whether the wake width of the irregular
plates scales according to the non-equilibrium dissipation law or
not. To do so, we assume that the average edges of the two wakes meet
where their wake extent $n\delta$ (with $n$ close to $2$) is equal to
$S/2$. Figure \ref{abb:sketch_wake_inter} visualizes this
configuration. Regarding this wake interaction process, we identify
the transition from regime 1 to regime 2 (defined as $x_{12}$) as the
beginning of significant interactions.

\begin{figure}
\centering
\subfigure[\label{abb_sigmaU_all}]{\includegraphics[width=0.45\textwidth]{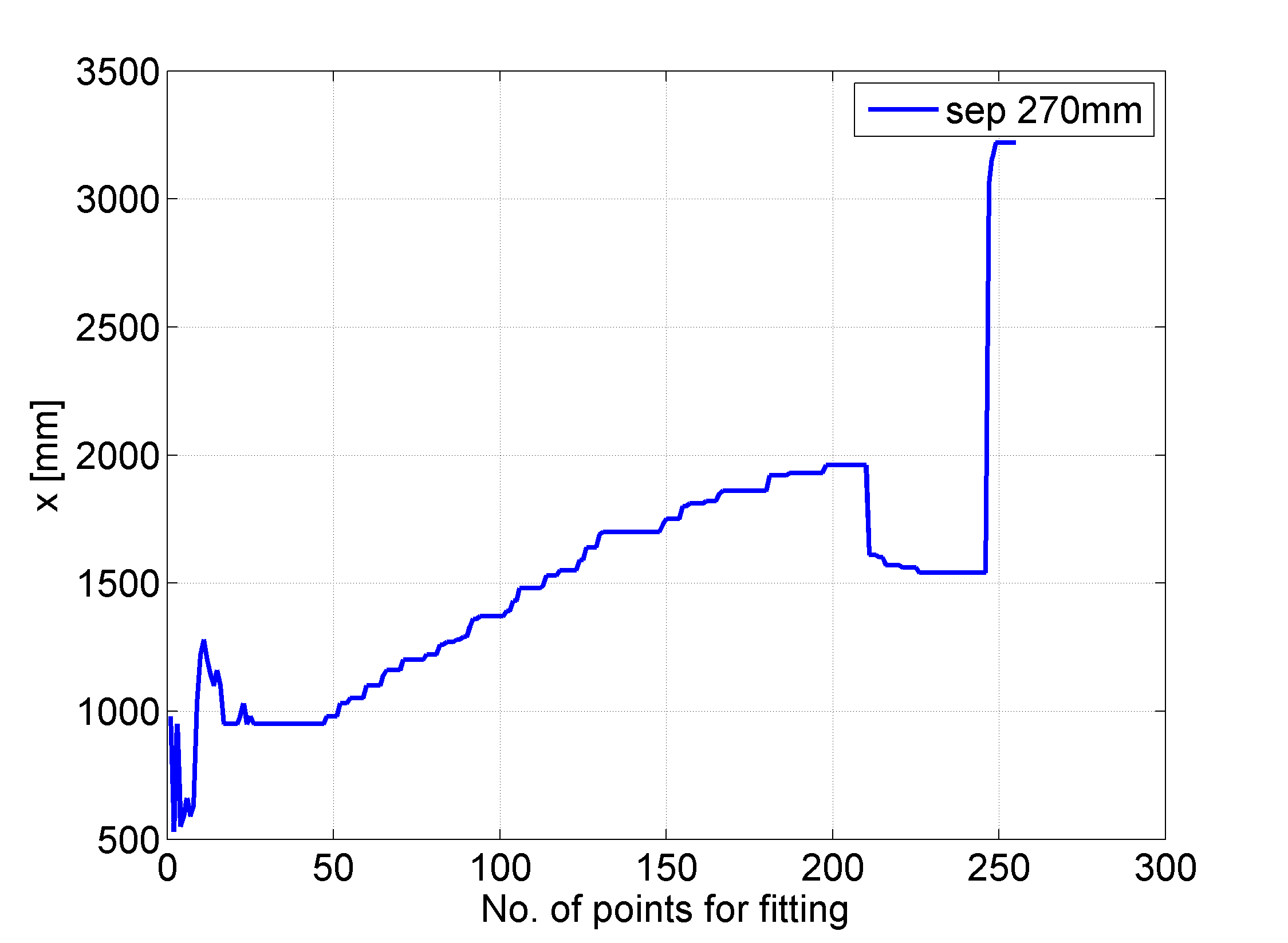}}
\subfigure[]{\includegraphics[width=0.45\textwidth]{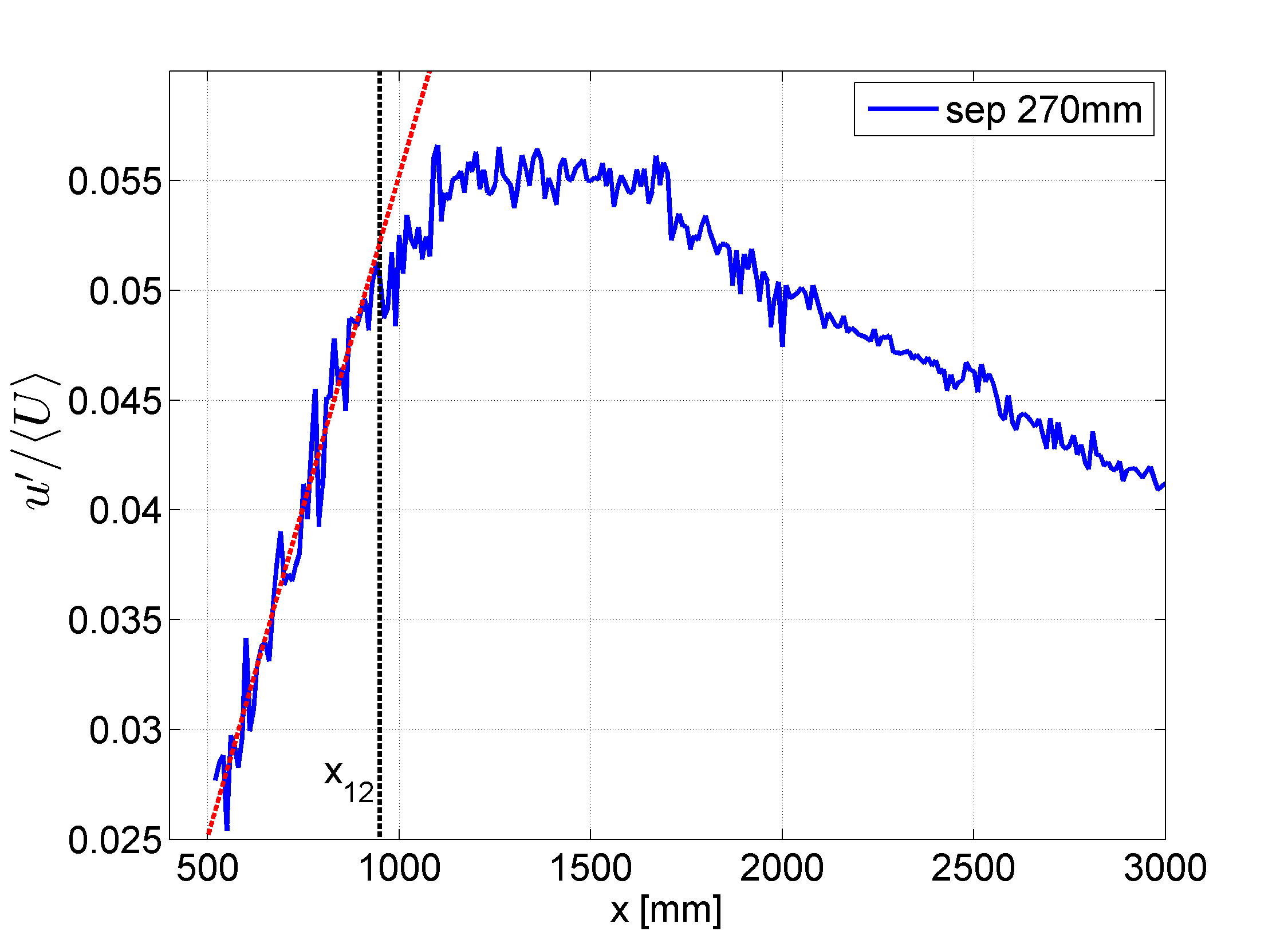}}
\caption{Example illustrating the procedure of automatic detection of
  $x_{12}$. Location of minimum quadratic deviation as a function of
  the number of points used for the linear fit (a). $x_{12}$
  determined and its corresponding linear fit (b).}
\label{abb_x12_seeking}
\end{figure}

Figure \ref{abb_x12_seeking}b shows an example of the streamwise
evolution of the normalised velocity fluctuations for the particular
case $S=270$mm (others are shown in figure 6a and present a similar
shape). Regarding the wake interaction process, we interpret the
transition point $x_{12}$ from region 1 to 2 along the centreline
streamwise axis to be the beginning of significant interactions
between the two wakes in terms of turbulence velocity fluctuations. It
can be observed that region 1 is characterised by a linear relation
between $u\prime/\langle U(x)\rangle$ and $x$. We identify the
streamwise transition point $x_{12}$ as the point where this linearity
is lost and use this property to determine $x_{12}$ automatically for
every case. To do so, the linear regime is fitted locally with
increasing number of points. Each linear fit is continuously
extrapolated on all points of the curve and the average quadratic
deviation is calculated. Plotting the location of lowest quadratic
deviation $x_{min}$ versus the number of points used for the linear
interpolation, the location $x_{12}$ can easily be detected by the
first plateau region within this plot (other plateau's may exist for
larger $x_{min}$ due to the change of concavity that occurs in regime
2). An example of this procedure is shown in figures
\ref{abb_x12_seeking}a\& b.

\begin{figure}
\centering
\includegraphics[width=0.95\textwidth]{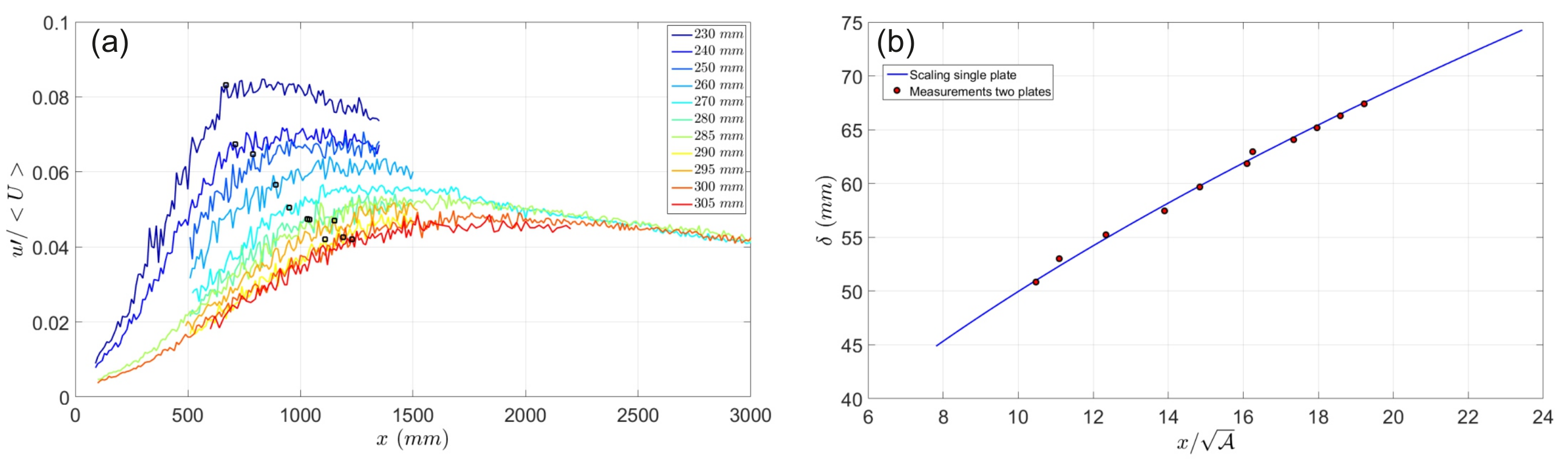}
\caption{(a) Streamwise profile of the velocity fluctuations for all
  plate separations $S$ in table \ref{tab_intersectionpoints}. The
  black squares represent $x_{12}$, i.e. the $x$-locations of the
  transition between regimes 1 and 2; (b) Comparison of the wake half
  width profile obtained for a single irregular plate
  by~\cite{dairay15} (blue line) with the results obtained by the
  wake-interaction setup used here (red line \& circles).}
\label{abb_wake_scaling}
\end{figure}

The streamwise locations found by this automatic procedure are
represented by the black squares in figure \ref{abb_wake_scaling}a. A
summary of the locations $x_{12}$ identified by this process is given
in table \ref{tab_intersectionpoints}. To determine the factor $n$, we
use the scaling law of the wake half width:
\begin{equation}
\frac{\delta(x)}{\theta}=B\left(\frac{x-x_0}{\theta}\right)^\beta
\label{eq:wakewidth}
\end{equation}
where $\delta(x)$ is the wake-width at position $x$, $\theta$ is
momentum thickness of the plate, $x_0$ stands for a virtual origin of
the plate, and $B$ and $\beta$ are fitting parameters. For the
irregular plates used herein, \cite{dairay15} found the following
values in agreement with the non-equilibrium dissipation law:
$\theta=21\,$mm, $x_0/\theta=-5.35$, $B=0.37$, and $\beta=0.52$. Using
these values and equation \ref{eq:wakewidth}, we can set up
equation \ref{eq:intersection} and solve it for $n$ for every plate
separation tested. The values thus obtained for $n$ are given in
table \ref{tab_intersectionpoints}.
\begin{equation}
\frac{1}{n}\cdot\frac{S}{2\,\theta}=B\left(\frac{x_{12}-x_0}{\theta}\right)^{\beta}
.
\label{eq:intersection}
\end{equation}
\begin{table}
\centering
\caption{Overview of transition locations between regime 1 \& 2\label{tab_intersectionpoints}}
\begin{tabular}{p{1.5cm}p{0.5cm}p{0.5cm}p{0.5cm}p{0.5cm}p{0.5cm}p{0.5cm}p{0.5cm}p{0.5cm}p{0.5cm}p{0.5cm}p{0.5cm}}
\hline\noalign{\smallskip}
$S$ [mm] & 230 & 240 & 250 & 260 & 270 & 280 & 285 & 290 & 295 & 300 & 305 \\
\noalign{\smallskip}\hline\noalign{\smallskip}
$x_{12}$ [mm] &  670 & 710 & 790 & 890 & 950 & 1030 & 1040 & 1110 & 1150 & 1190 & 1230\\
\noalign{\smallskip}\hline\noalign{\smallskip}
$n$ &  2.26 & 2.29 & 2.28 & 2.24 & 2.26 & 2.26 & 2.29 & 2.26 & 2.26 & 2.26 & 2.26 \\
$\langle n\rangle$ & 2.26 & & & & & & & & & &\\
$\sigma_n/\langle n\rangle$ & 0.0068 & & & & & & & & & &\\
\noalign{\smallskip}\hline\noalign{\smallskip}
\end{tabular}
\end{table}  
Perhaps remarkably, all $n$-factors are found to be close to the mean
value $\langle n\rangle=2.26$. This result indicates that significant
wake interactions take place where $2.26\cdot\delta(x)$ is equal to
half the plate separation. Moreover, this result is supported by
previous findings of~\cite{dairay15} who conducted vertical velocity
profiles at distinct locations downstream of an irregular plate and
found self-similar profiles for mean velocity as well as streamwise
fluctuations: the radial location $2.26\,\delta$ coincides well with
relatively high levels of turbulence fluctuations and relatively high
cross-stream gradients of the mean streamwise velocity.

This result supports the validity of the non-equilibrium dissipation
law in the wake of the irregular plates and determines the beginning
of wake interactions on the basis of the non-equilibrium scalings of
the wake half width. To underscore this result, figure
\ref{abb_wake_scaling}b shows the streamwise evolution of the wake
half width with increasing streamwise position from the plate as
obtained for a single plate by~\cite{dairay15} (blue line). To compare
it with the result we just obtained with our two wake interaction
set-up, we calculate $\delta_{wake}(x)=S/(2\,\langle n\rangle)
\left(\frac{x-x_0}{x_{12}-x_{0}}\right)^{0.52}$ for $x$ larger than
$x_{12}$ and plot it for each plate separation against $x$
normalised by the reference length $L_b=\sqrt{\cal{A}}=64$~mm. Again,
the results coincide. It is therefore possible to predict the overall
level of the streamwise profile of $\delta (x)$, which is a
characteristic of the mean flow cross-stream profile of a single wake,
by acquiring velocity fluctuation measurements in regions 1 and 2 of a
two-wake set-up.

\subsection{Higher-Order Moments}

\begin{figure}
\centering
\subfigure[\label{abb_all_skew}]{\includegraphics[width=0.45\textwidth]{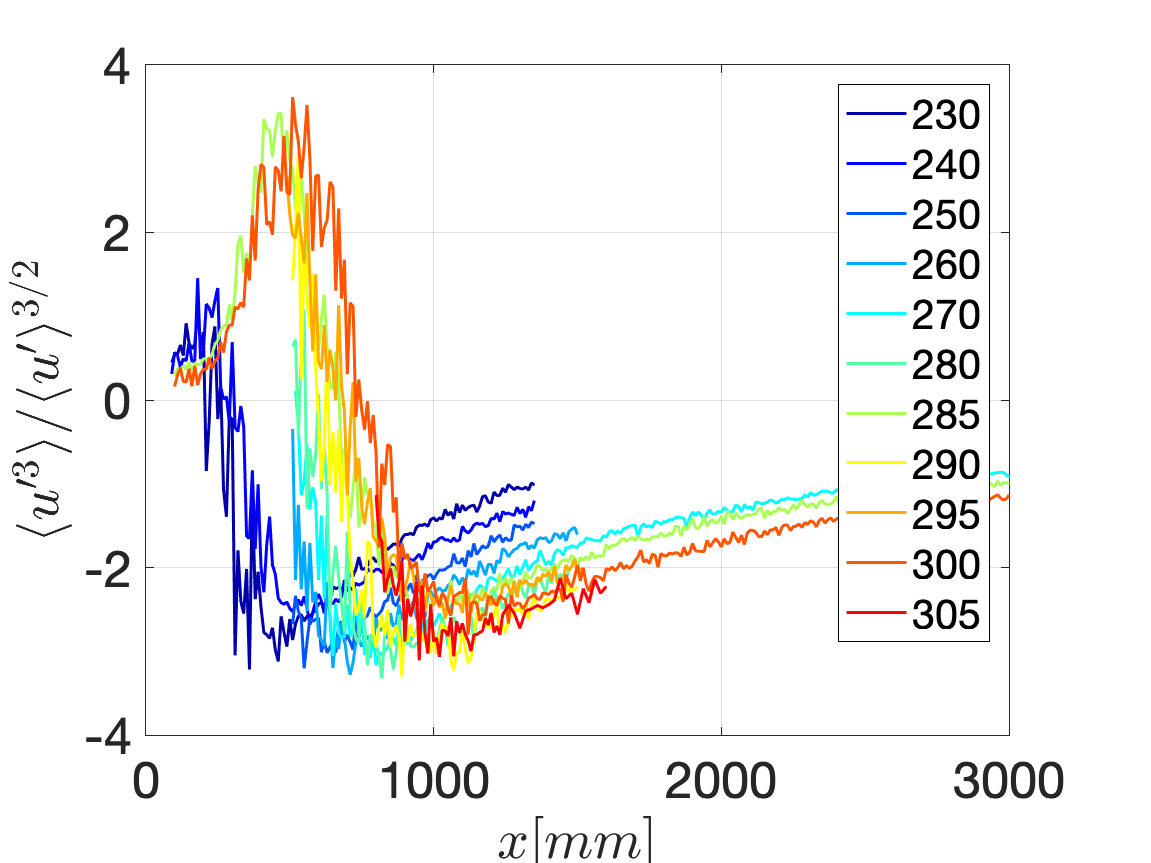}}
\subfigure[\label{abb_all_flat}]{\includegraphics[width=0.45\textwidth]{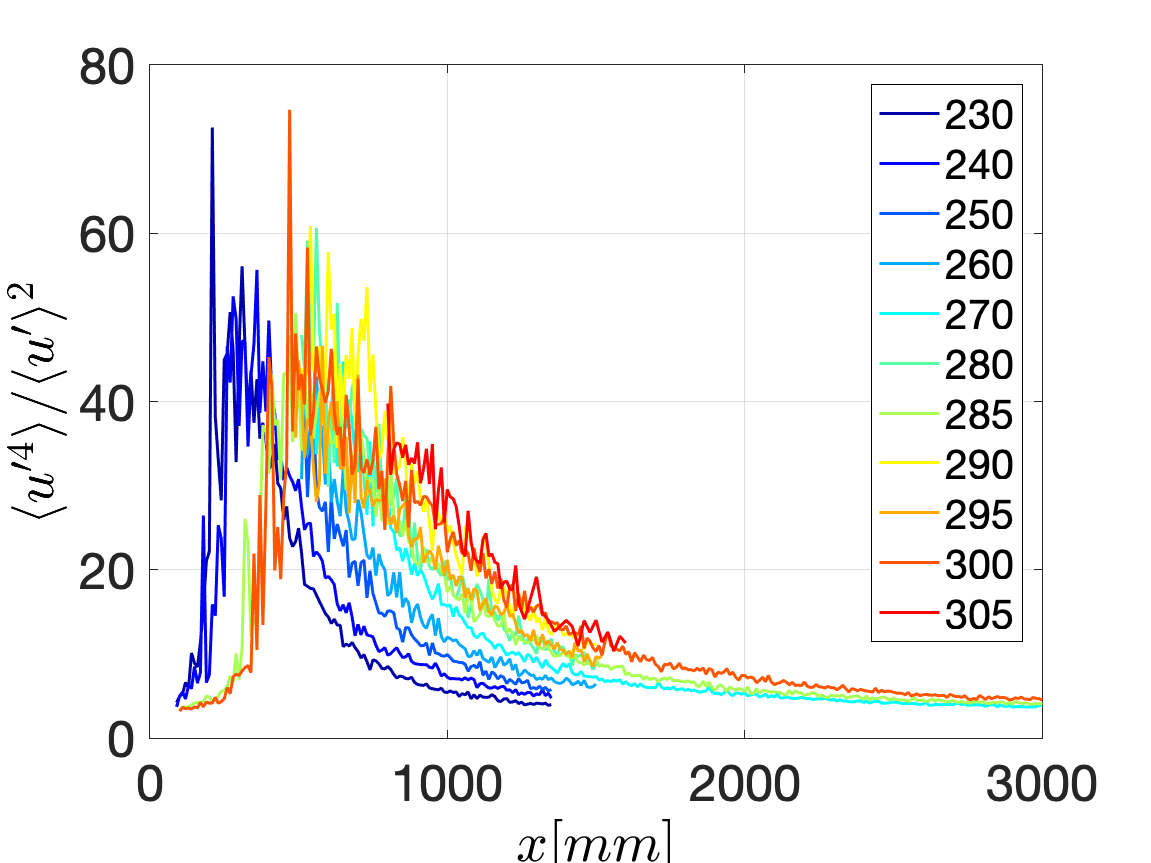}}
\caption{(a) Skewness and (b) flatness profiles for all plate separations tested. The separation increases from left to right.}
\end{figure}

In previous studies, wake interactions have been investigated in terms
of higher-order moments, such as velocity skewness and
flatness~(\cite{Mazellier2010,Laizet2015}). Figure \ref{abb_all_skew}
and \ref{abb_all_flat} show the skewness and flatness profiles for all
plate separations tested. The behavior of both moments is quite
similar to the results of~\cite{Mazellier2010,Laizet2015}. Consider
that these two studies both investigated 2D, planar wakes of the bars
of turbulence grids. In contrast, the wakes in the current study are
3D and axisymmetric. Hence, the similarity of the high-order moments
is non-trivial. 
The detailed analysis of this finding is left for future studies.

\begin{figure}
\centering
\subfigure[\label{abb_all_skew_scaled}]{\includegraphics[width=0.45\textwidth]{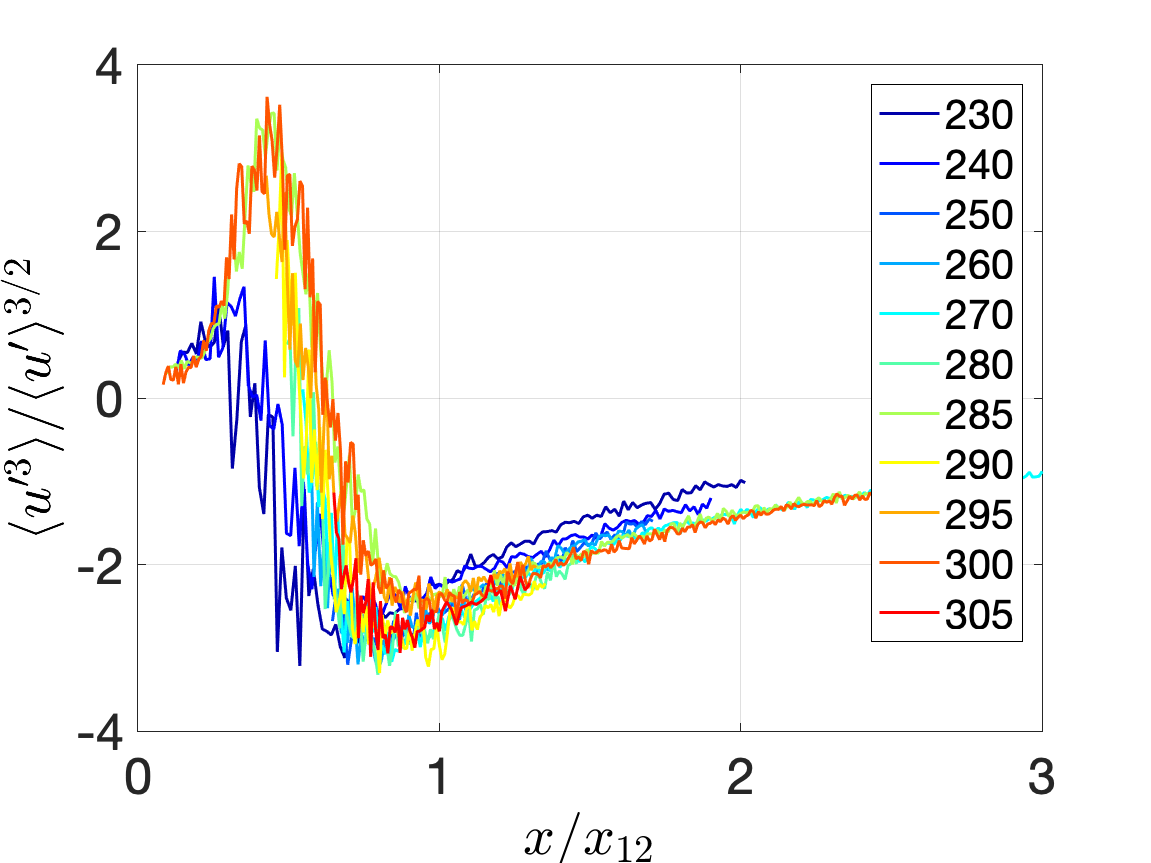}}
\subfigure[\label{abb_all_flat_scaled0}]{\includegraphics[width=0.45\textwidth]{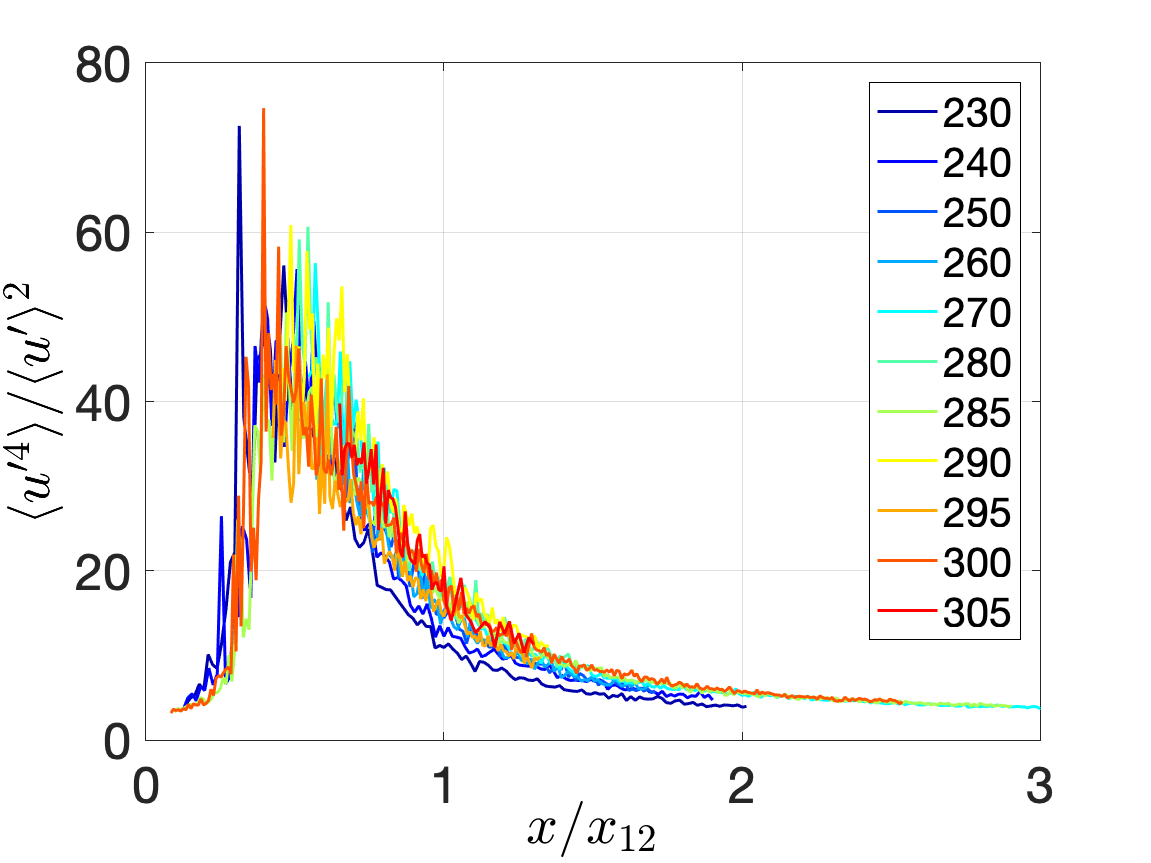}}
\caption{(a) Skewness and (b) flatness profiles for all plate separations tested. The streamwise position $x		$ is scaled by the length-scale $x_{12}$.}
\end{figure}

\cite{Mazellier2010} showed that the skewness and flatness profiles of
different turbulence grids could be collapsed by scaling the
streamwise position with a wake-interaction length scale
$x_*$. Subsequently, \cite{gomes2012} collapsed velocity fluctuation
profiles of different experiments with an improved wake-interaction
length scale $x'_*$. In figures \ref{abb_all_skew_scaled} and
\ref{abb_all_flat_scaled0} the skewness and flatness profiles are replotted
with the $x$-axis scaled by the length-scale $x_{12}$ discussed
earlier. We find that all values of $S$ collapse, with the exception
of the smallest separation; $S=230mm$. The reason for this exception
may be that the wakes meet before having become self-similar or
axisymmetric for this low value of $S$.

In the following sub-section, we further test the idea that the
length-scale $x_{12}$ is a wake-interaction length-scale and show also
that it can be used
to scale the streamwise profile of the turbulent velocity
fluctuations.

\subsection{Wake-Interaction Length Scale}

As stated above, \cite{Mazellier2010} and \cite{gomes2012} proposed
formulae for a wake-interaction length scale characterising the
interaction of planar wakes emanating from fractal and regular
grids. This length-scale is easily constructed for any free-shear
flow. In our case, two axisymmetric wakes can be surely expected to
interact at a streamwise distance $x^*$ from the plates where $\delta$
is equal to half the distance between the centres of the plates
(defined as $S$ in this work). This implies that the wake interaction
length $x^*$ scales as $x^* \propto S^2$ on the basis of equation \ref{eq2},
resulting in
\begin{equation}
 \left( \frac{x^{*}-x_0}{\theta} \right) \propto \left(
 \frac{S}{2\theta} \right)^2.
\label{eq_link}
\end{equation}
A crucial step in the wake interaction approach of~\cite{gomes2012} is
the neglect of the virtual origin $x_0$ because the \textit{``distance
  downstream where the wakes meet is very much larger than the virtual
  origin. We therefore ignore the virtual origin $x_0$ and
  effectively set it equal to zero.''} However, \cite{Nedic2013a}
showed that simply setting the virtual origin to zero is not valid for
centreline measurements of axisymmetric wakes and can yield false
results. Nevertheless, since the current measurements are conducted
between 1.8 and 2.7 characteristic lengths off the centerlines of the
wakes, the influence of the virtual origin $x_0$ is indeed
negligible. To verify this, we consider the derivative of the velocity
deficit with respect to the virtual origin
$\partial(u_0/U_\infty)/\partial(x_0/\theta)$. According
to~\cite{cafiero2020length}, the evolution of the velocity deficit in
the streamwise direction $x$ and the radial direction $y$ for both
sets of plates in the relevant streamwise range is,

\begin{equation}
\frac{u_0}{U_\infty} = A\left( \frac{x-x_0}{\theta} \right)^\alpha e^{\left(-\frac{ay^2}{\delta^2}\right)}.\label{eq_velodef}
\end{equation}
\noindent
The diferentiation yields
\begin{align}
\frac{\partial\frac{u_0}{U_\infty}}{\partial \frac{x_0}{\theta}} =& -A\alpha\left(\frac{x-x_0}{\theta}\right)^{\alpha-1}e^{\left(-\frac{ay^2}{\delta^2}\right)} \nonumber\\
&+ A\alpha\left(\frac{x-x_0}{\theta}\right)^{\alpha}e^{\left(-\frac{ay^2}{\delta^2}\right)}\left[ -2a\beta y^2(B\theta)^{-2}\left(\frac{x-x_0}{\theta} \right)^{-2\beta-1} \right].
\end{align}
Setting the non-equilibrium values of the scaling exponents
$\alpha=-1$ and $\beta=1/2$ (see ~\cite{dairay15}), the result is that
\begin{equation}
\frac{\partial\frac{u_0}{U\infty}}{\partial \frac{x_0}{\theta}} =
-A\left(\frac{x-x_0}{\theta}\right)^{-2}\underbrace{e^{\left(-\frac{ay^2}{\delta^2}\right)}\left[1-\frac{ay^2}{\delta^2}\right]}_{\text{radial
    dependency }f(y/\delta)}.
\label{eq_deri}
\end{equation}
\begin{figure}
\centering
\includegraphics[width=0.6\textwidth]{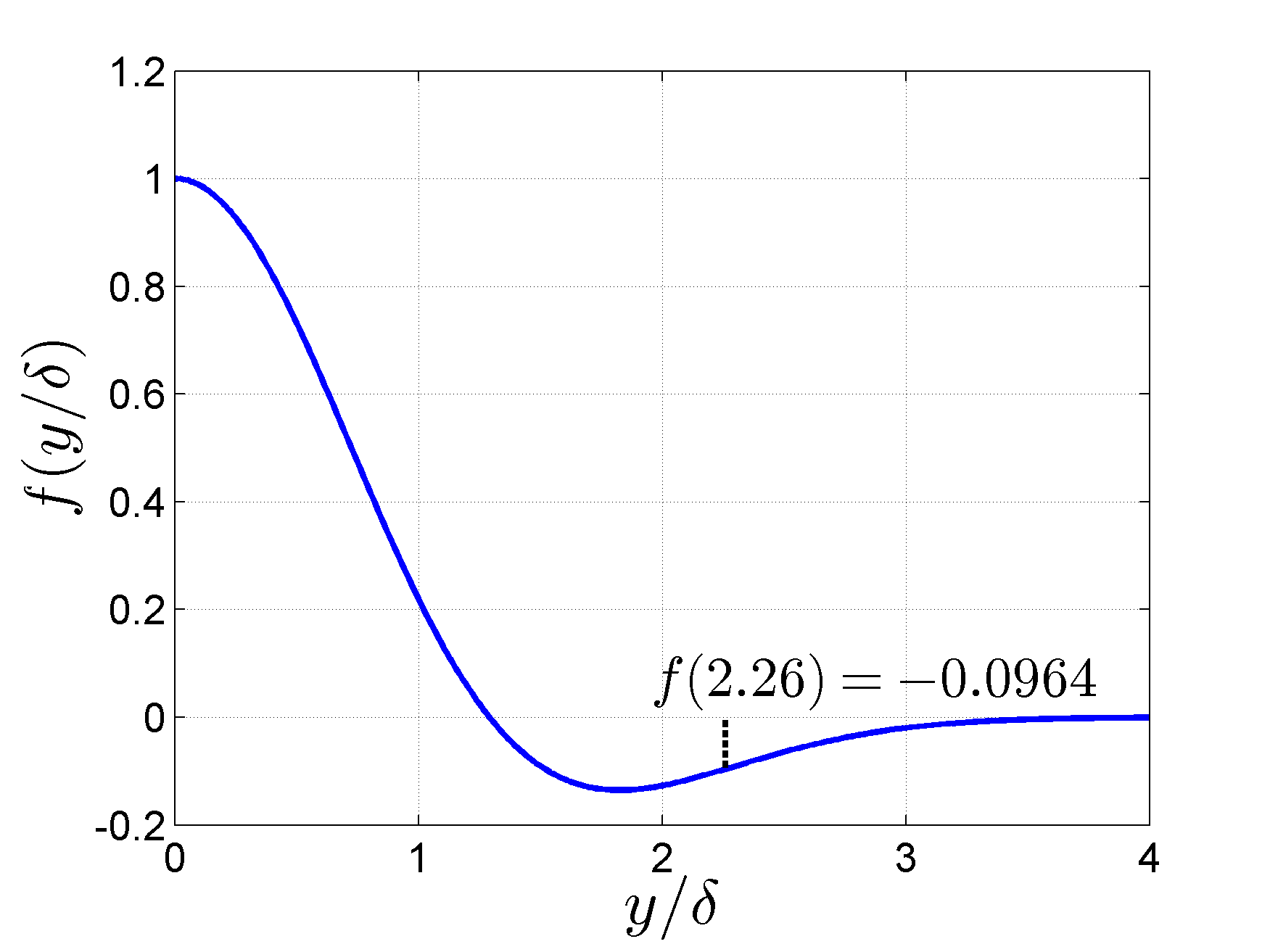}
\caption{Radial distribution of $\partial(u_0/U_\infty)/\partial(x_0/\theta)$.}
\label{abb_1}
\end{figure}
The function $f(y/\delta)$ determines the radial distribution of the
derivative $\partial(u_0/U_\infty)/\partial(x_0/\theta)$. A plot of
$f(y/\delta)$ is given in figure \ref{abb_1}, with
$a=0.6$~(\cite{Nedic2013}). As mentioned above, we identify the
meeting of the two plate wakes at the radial location
$y=2.26\delta$. At this radial location, the modulus of the derivative
has dropped to approximately $9.6\%$ of its centerline value. Hence,
the influence of the virtual origin $x_0/\theta$ on the velocity
deficit is marginal for small variations and can be neglected. We can
then define a modified wake interaction length for two axisymmetric
plates in the non-equilibrium regime by,

\begin{equation}
x_{*}' = \phi \frac{S^2}{4 \theta}.
\label{WIL}
\end{equation}

\noindent where $\phi$ is a constant that depends on the geometry of
the plates and accounts for the contributions of $B$ and $\langle
n\rangle$.

\begin{table}
\centering
\caption{Wake-interaction length scale\label{tab_wake_interscale}}
\begin{tabular}{p{1.5cm}p{0.5cm}p{0.5cm}p{0.5cm}p{0.5cm}p{0.5cm}p{0.5cm}p{0.5cm}p{0.5cm}p{0.5cm}p{0.5cm}p{0.5cm}}
\hline\noalign{\smallskip}
$S$ [mm] & 230 & 240 & 250 & 260 & 270 & 280 & 285 & 290 & 295 & 300 & 305 \\
$x_{12}$ [mm] &  670 & 710 & 790 & 890 & 950 & 1030 & 1040 & 1110 & 1150 & 1190 & 1230\\
$x_*'$ [mm] & 630 & 686 & 744 & 805 & 868 & 933 & 967 & 1001 & 1036 & 1071 & 1107 \\
$x_*'/x_{12}$ & 0.94 & 0.97 & 0.94 & 0.90 & 0.91 & 0.91 & 0.93 & 0.90 & 0.90 & 0.90 & 0.90 \\
\noalign{\smallskip}\hline\noalign{\smallskip}
\end{tabular}
\vspace{-10pt}
\end{table}  
The wake-interaction length scale $x_*'$ based on equation \ref{WIL}
is given in table \ref{tab_wake_interscale} for different values of
$S$ together with the length scale $x_{12}$ determined earlier. It can
be see that $x_{12}$ and $x_*'$ are proportional to each other.
Hence, the length scale $x_{12}$, determined empirically, scales with
the modified wake-interaction length scale $x_*'$ based on the
non-equilibrium dissipation law. It is therefore possible to consider
$x_{12}$ as the wake interaction length for turbulent axisymmetric
wakes. As a result, the length scale $x_{12}$ can be used to collapse
the streamwise development of the turbulence intensities. To complete
the scaling of the turbulence intensity we note that, in the case of a
single wake, the turbulent kinetic energy and the Reynolds shear stress
evolve together in the streamwise direction (\cite{townsend80,george89,
dairay15}) and that the Reynolds shear stress scales
with streamwise distance as $U_\infty u_0(d/dx\,\delta(x))$ (see
\cite{george89,dairay15}). We therefore attempt the scaling
$u\prime \sim \sqrt{U_\infty u_0(d/dx\,\delta(x))} F(x/x_{12})$ where
$F$ is a dimensionless function of $x/x_{12}$. In this work we did not
measure the values of $u_0(x)$ and $\delta(x)$ and we therefore used
their fits from~\cite{dairay15} for the irregular plates.

\begin{figure}
\centering
\includegraphics[width=0.5\textwidth]{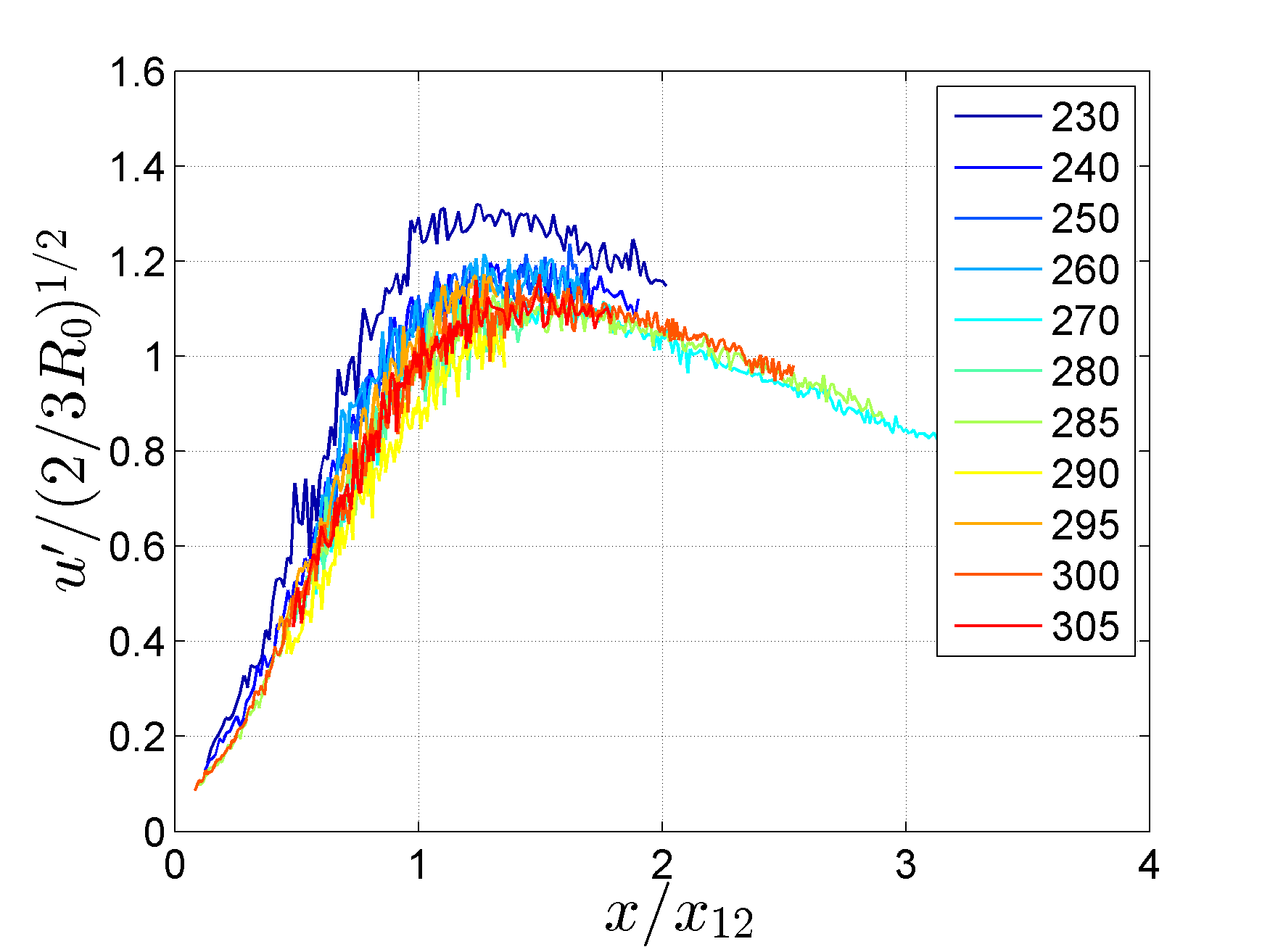}
\caption{Profiles of $u\prime$ obtained for plates with irregular
  peripheries normalised according to according to equation $u\prime
  \sim \sqrt{2/3 R_{0}} F(x/x_{12})$ where $R_0 = U_\infty
  u_0(d/dx\,\delta(x))$.}
\label{abb_uprime_scaled}
\end{figure}

In figure \ref{abb_uprime_scaled} we show the velocity fluctuations
normalised according to $u\prime \sim \sqrt{U_\infty
  u_0(d/dx\,\delta(x))} F(x/x_{12})$. Most of the profiles collapse
fairly well. Consistently with the discussion in sub-section 3.2, the
collapse is worse for smaller values of $S$.

%
\section{Results - Regular Square Plates}

\begin{figure}
\centering
\subfigure[\label{abb_square_all_uprime}]{\includegraphics[width=0.45\textwidth]{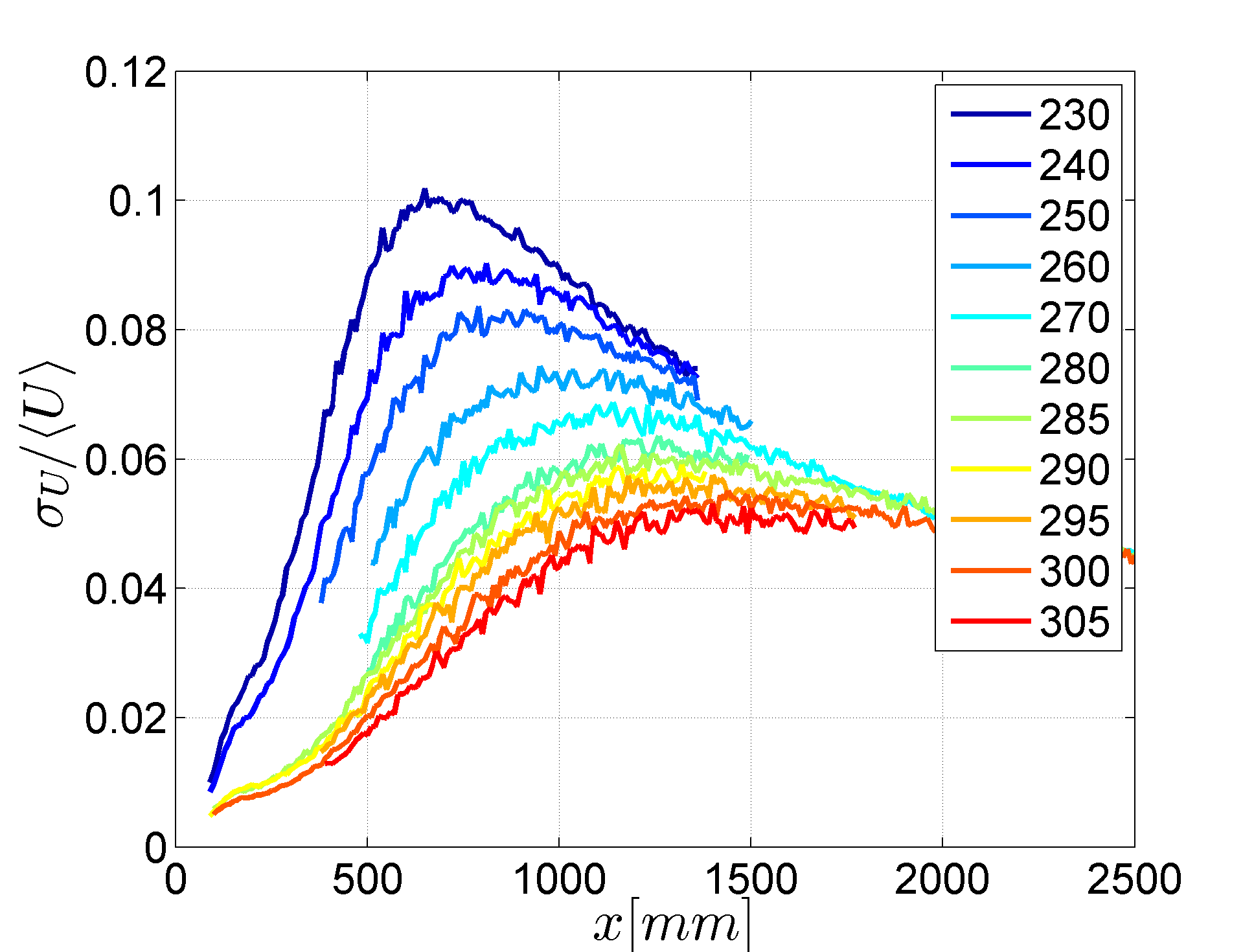}}
\subfigure[\label{abb_all_flat_scaled}]{\includegraphics[width=0.45\textwidth]{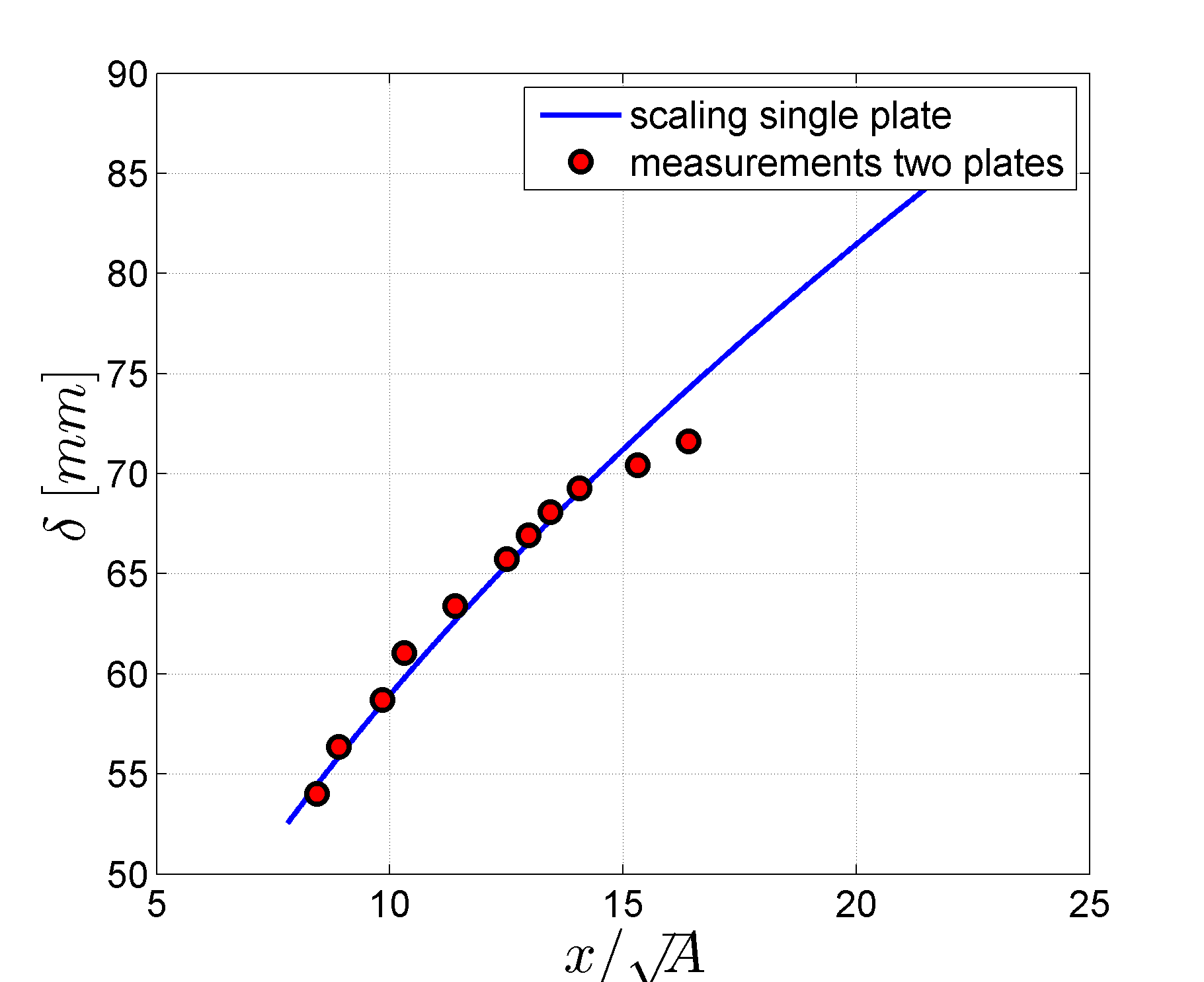}}
\caption{(a) Streamwise distribution of the velocity fluctuations for
  all square plate's separations given in table
  \ref{tab_intersectionpoints}. (b) Comparison of the wake-width
  distribution found for a single regular plate by~\cite{obligado2015}
  (blue line) and the results gained by the wake-interaction setup
  used herein (red line \& circles). The parameters of equation
  \ref{eq:wakewidth} corresponding to square plates are
  $\theta=19.7\,$mm, $x_0/\theta=-6.20$, $B=0.494$, and $\beta=0.50$
  (\cite{obligado2015}).}
\end{figure}

In addition to the two plates with irregular peripheries, two regular
square plates with the same area ($\sqrt{\cal{A}}=64$~mm) were also
investigated. Again eleven different separations were set and the same
measurements as for the irregular plates were conducted. Figure
\ref{abb_square_all_uprime} shows the streamwise development of the
velocity fluctuations for all separations. The shapes of these
profiles are similar to those for the irregular plates.
%
%
\\\indent Again, we determine the length-scale $x_{12}$ at the point
of transition from regime 1 to regime 2 in the turbulent fluctuations
streamwise profile (see figure \ref{abb_square_all_uprime}). The
results of this automated procedure are given in table
\ref{tab_results_squares}. The n-factor is once again very similar for
all values of $S$, but slightly different from the irregular plates:
$\langle n\rangle=2.13$. Figure \ref{abb_all_flat_scaled} compares the
wake-width based on the identified length-scale $x_{12}$ with the
results of the single-plate measurements of~\cite{obligado2015}.
Equation \ref{WIL} yields the modified wake-interaction length scale
$x_*'$ for these regular square plates (see table
\ref{tab_results_squares}). As found for the irregular plates, our
length-scale $x_{12}$ is a multiple of $x_*'$, in this regular case
with $x_{12}=1.24 \cdot x_*'$.
\begin{table}
\centering
\caption{Overview of the length-scale $x_{12}$ and other parameters
  chacraterizing the wake's interaction for the square
  plates\label{tab_results_squares}}
\begin{tabular}{p{1.8cm}p{0.5cm}p{0.5cm}p{0.5cm}p{0.5cm}p{0.5cm}p{0.5cm}p{0.5cm}p{0.5cm}p{0.5cm}p{0.5cm}p{0.5cm}}
\hline\noalign{\smallskip}
$S$ [mm] & 230 & 240 & 250 & 260 & 270 & 280 & 285 & 290 & 295 & 300 & 305 \\
$x_{12}$ [mm] & 540 & 570 & 630 & 660 & 730 & 800 & 830 & 860 & 900 & 980 & 1050 \\
$n$ &  2.11 & 2.15 & 2.14 & 2.17 & 2.15 & 2.14 & 2.14 & 2.14 & 2.14 & 2.09 & 2.05 \\
$\langle n\rangle$ & 2.13 &  & & & & & & & & & \\
$\sigma_n/\langle n\rangle$ & 0.016 &  & & & & & & & & & \\
$x_*'$ [mm] & 671 & 731 & 793 & 858 & 925 & 995 & 1031 & 1067 & 1104 & 1142 & 1181 \\
$x_*'/x_{12}$ & 1.24 & 1.28 & 1.26 & 1.30 & 1.27 & 1.24 & 1.24 & 1.24 & 1.23 & 1.17 & 1.12 \\
\hline\noalign{\smallskip}
\end{tabular}
\vspace{-10pt}
\end{table}  
\\

\begin{figure}
\centering
\subfigure[\label{abb_23_skew_square}]{\includegraphics[width=0.45\textwidth]{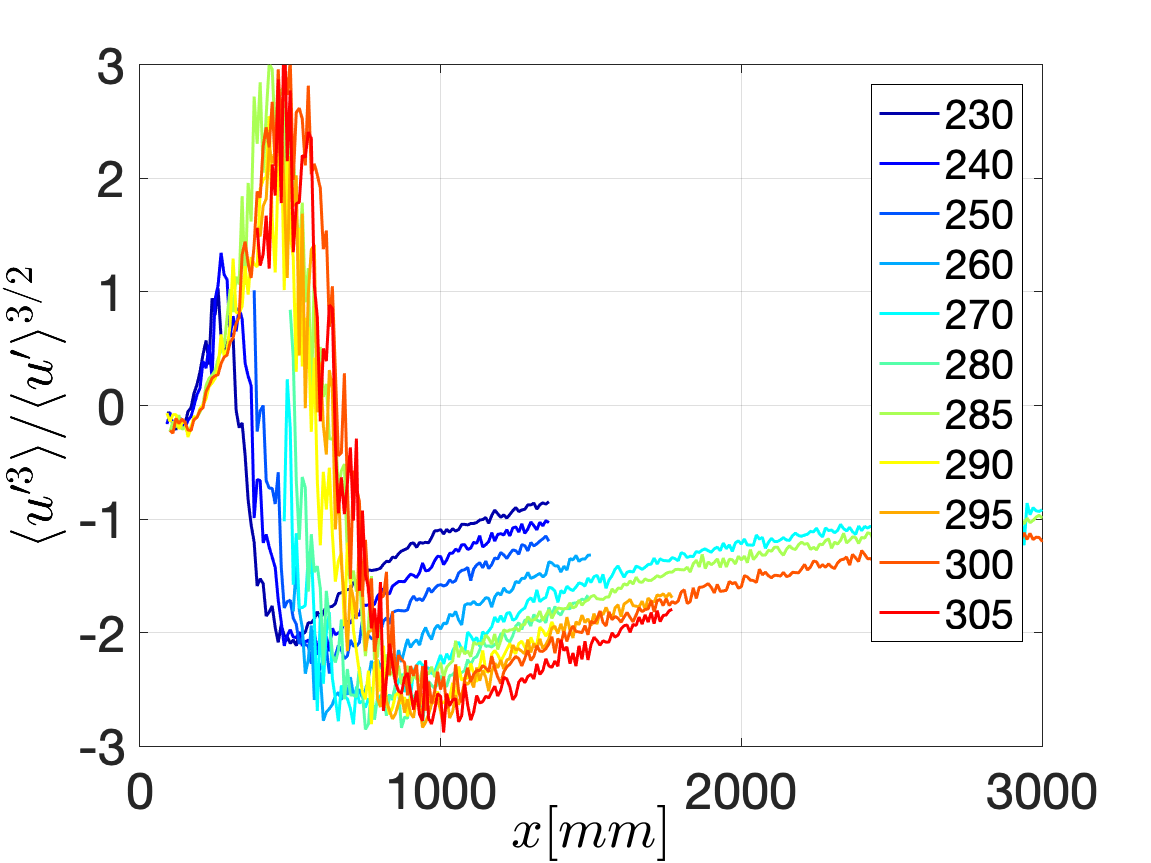}}
\subfigure[\label{abb_23_flat_square}]{\includegraphics[width=0.45\textwidth]{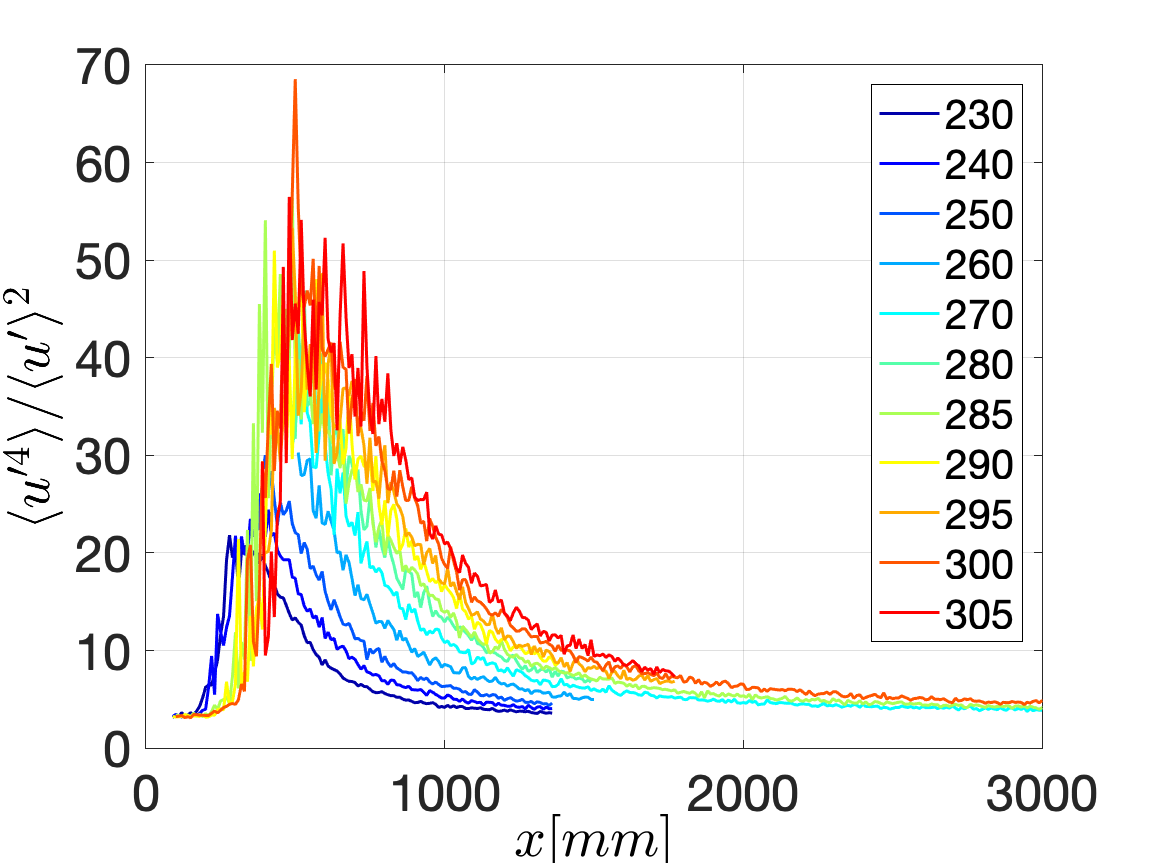}}\\
\subfigure[\label{abb_24_skew_square_scaled}]{\includegraphics[width=0.45\textwidth]{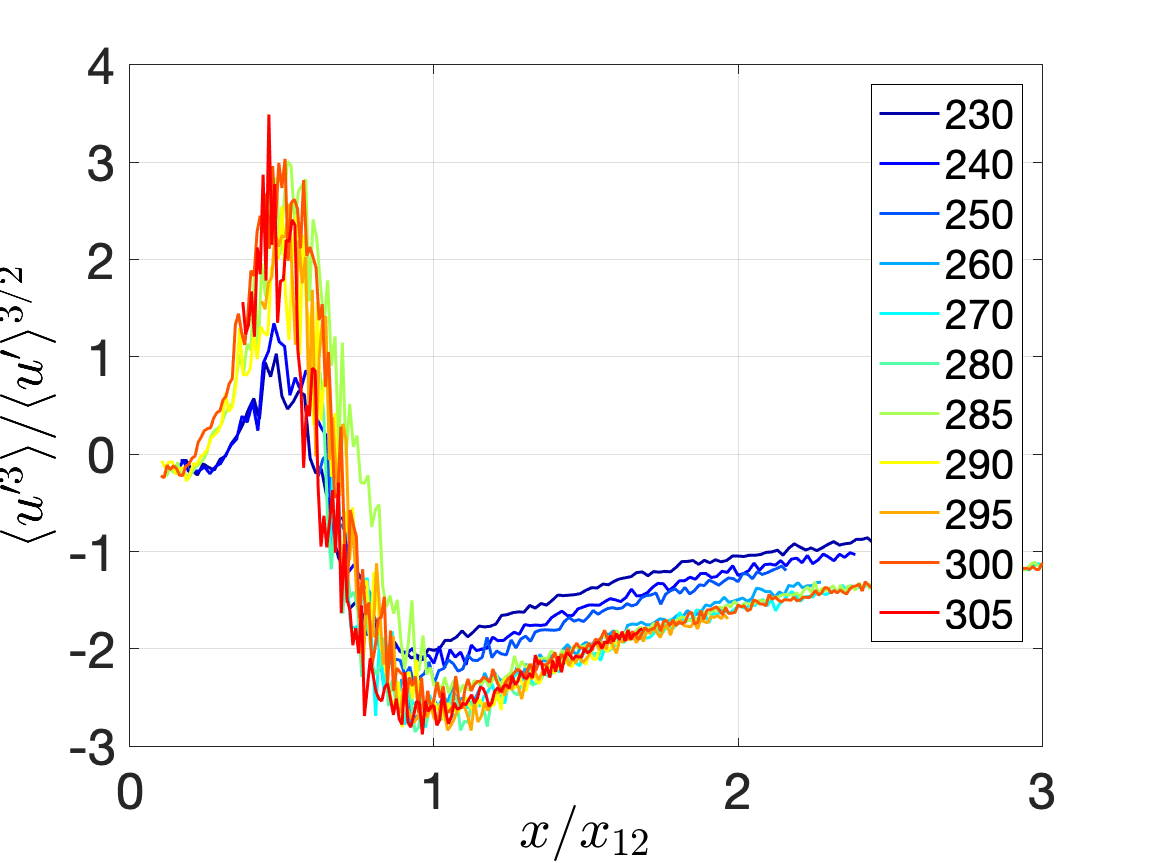}}
\subfigure[\label{abb_24_flat_square_scaled}]{\includegraphics[width=0.45\textwidth]{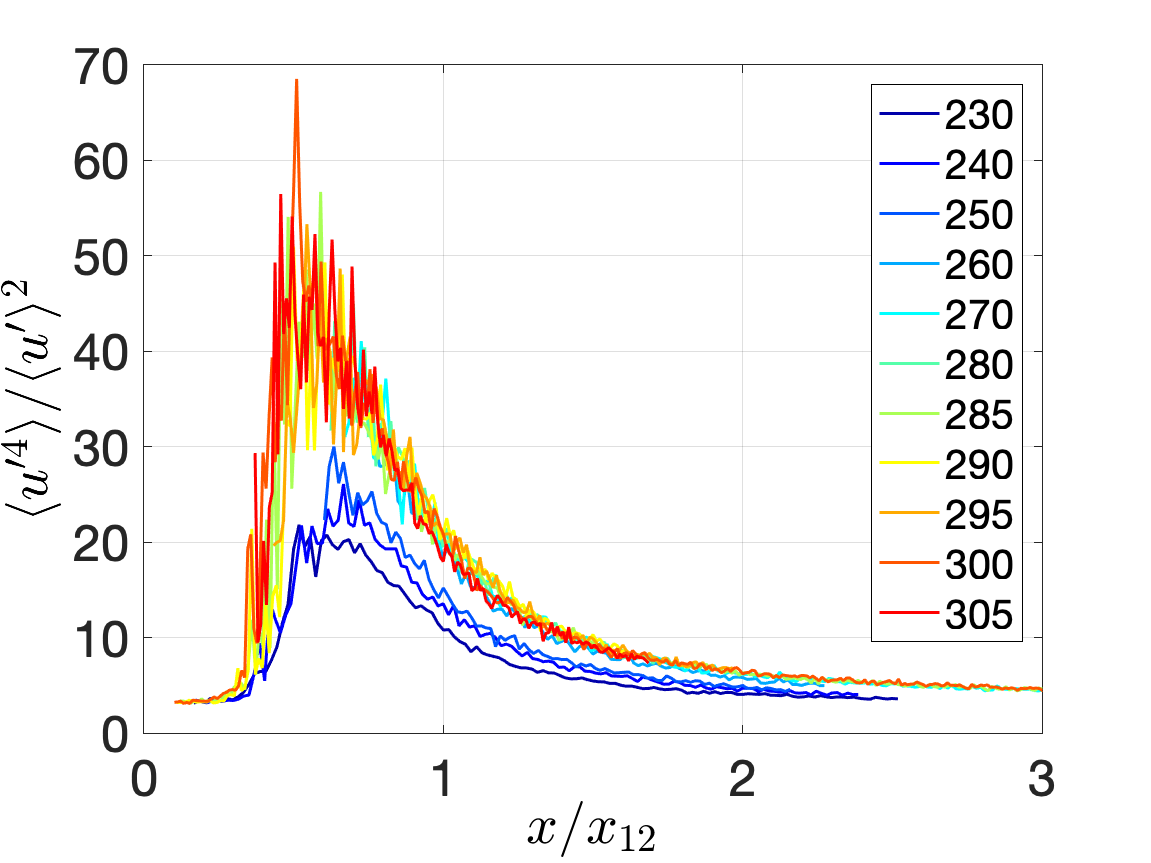}}
\caption{(a, b) Unscaled skewness and flatness profiles for all square plate separations tested. (c, d) Skewness and flatness profiles scaled with $x_{12}$.}
\end{figure}

\begin{figure}
\centering
\includegraphics[width=0.5\textwidth]{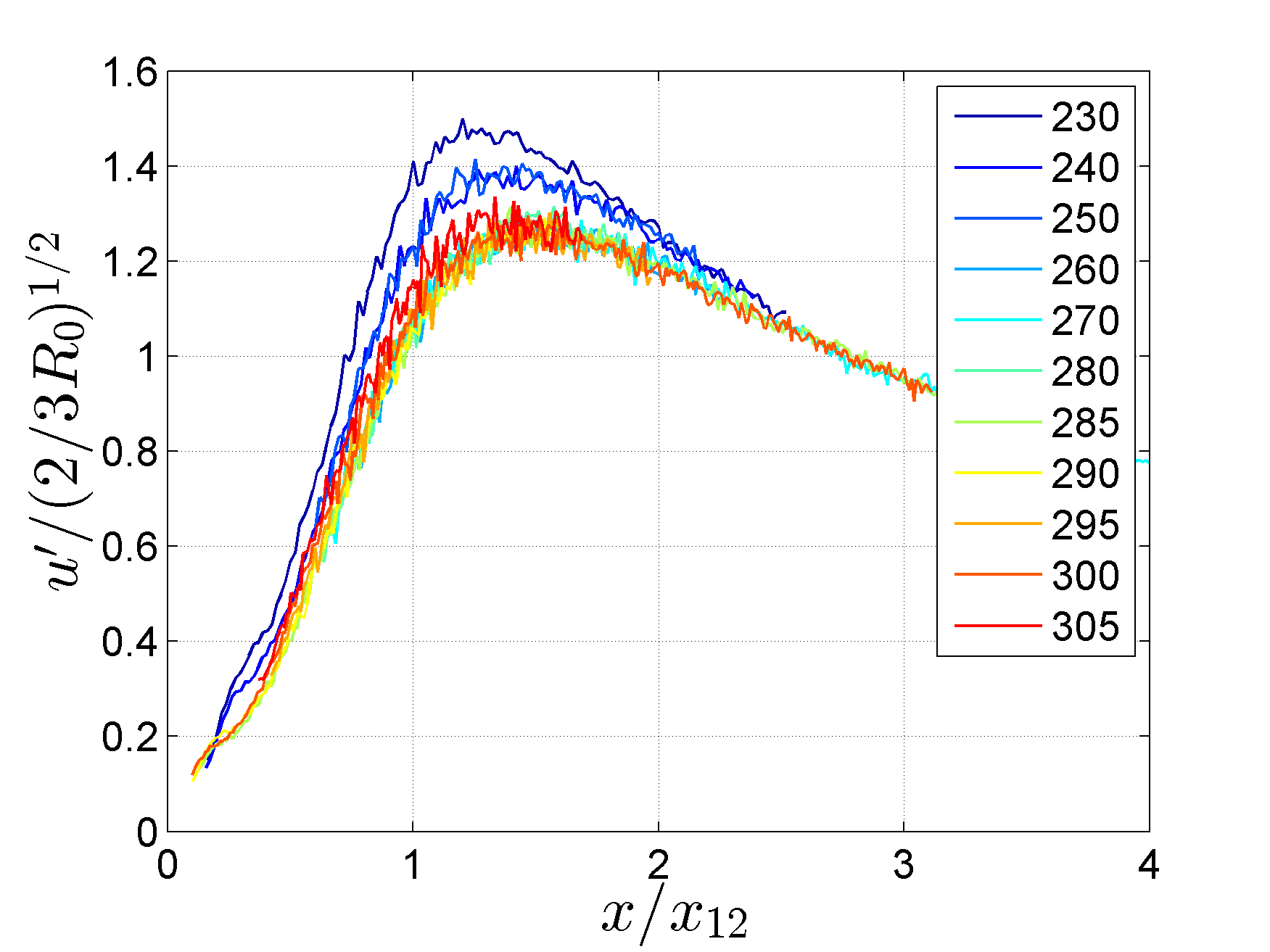}
\caption{Profiles of $u\prime$ obtained for plates with square
  peripheries normalised according to equation $u\prime \sim \sqrt{2/3
    R_{0} } F(x/x_{12})$ where $R_0 = U_\infty u_0(d/dx\,\delta(x))$.}
\label{abb_25_square_uprime_scaled}
\end{figure}

\indent To evaluate the collapse capability of $x_{12}$, we scale the
streamwise profile of the higher-order moments (skewness and flatness)
of the velocity fluctuations with $x_{12}$. In figures
\ref{abb_23_skew_square} and \ref{abb_23_flat_square} we plot the
original skewness and flatness profiles and in figures
\ref{abb_24_skew_square_scaled} and \ref{abb_24_flat_square_scaled} we
plot the streamwise skewness and flatness profiles scaled with
$x_{12}$. Some collapse, particularly for the larger values of $S$ is,
once again, observed.
\\ \indent
Finally, in figure \ref{abb_25_square_uprime_scaled} $x_{12}$ is used
to collapse the streamwise profile of velocity
fluctuations. Streamwise velocity fluctuations are scaled according to
$u\prime \sim \sqrt{U_\infty u_0(d/dx\,\delta(x))} F(x/x_{12})$ as for
the irregular plates. A fairly good collapse of the profiles is
obtained again except for the smaller values of $S$.
\clearpage
\section{Results - Comparison irregular and regular plates}
\begin{figure}
\centering
\subfigure[\label{abb_24_ALL_SKEW}]{\includegraphics[width=0.45\textwidth]{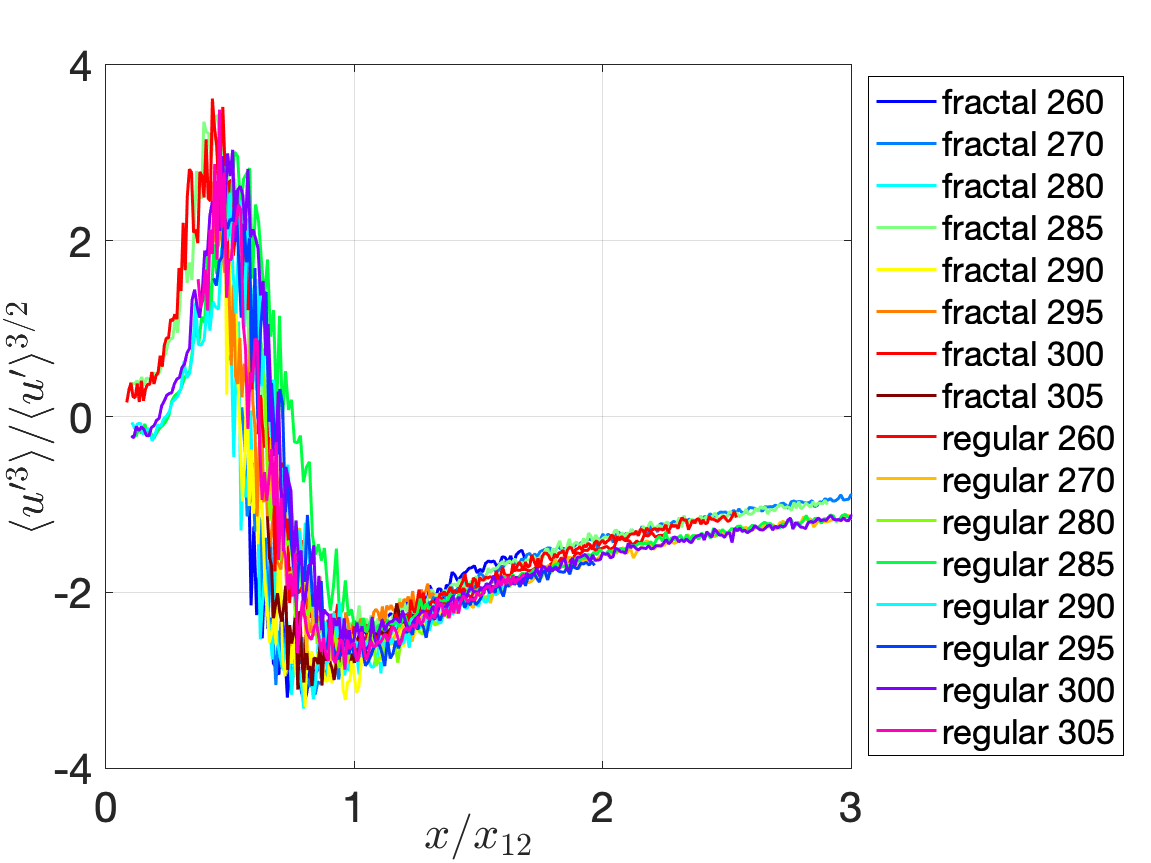}}
\subfigure[\label{abb_24_ALL_FLAT}]{\includegraphics[width=0.45\textwidth]{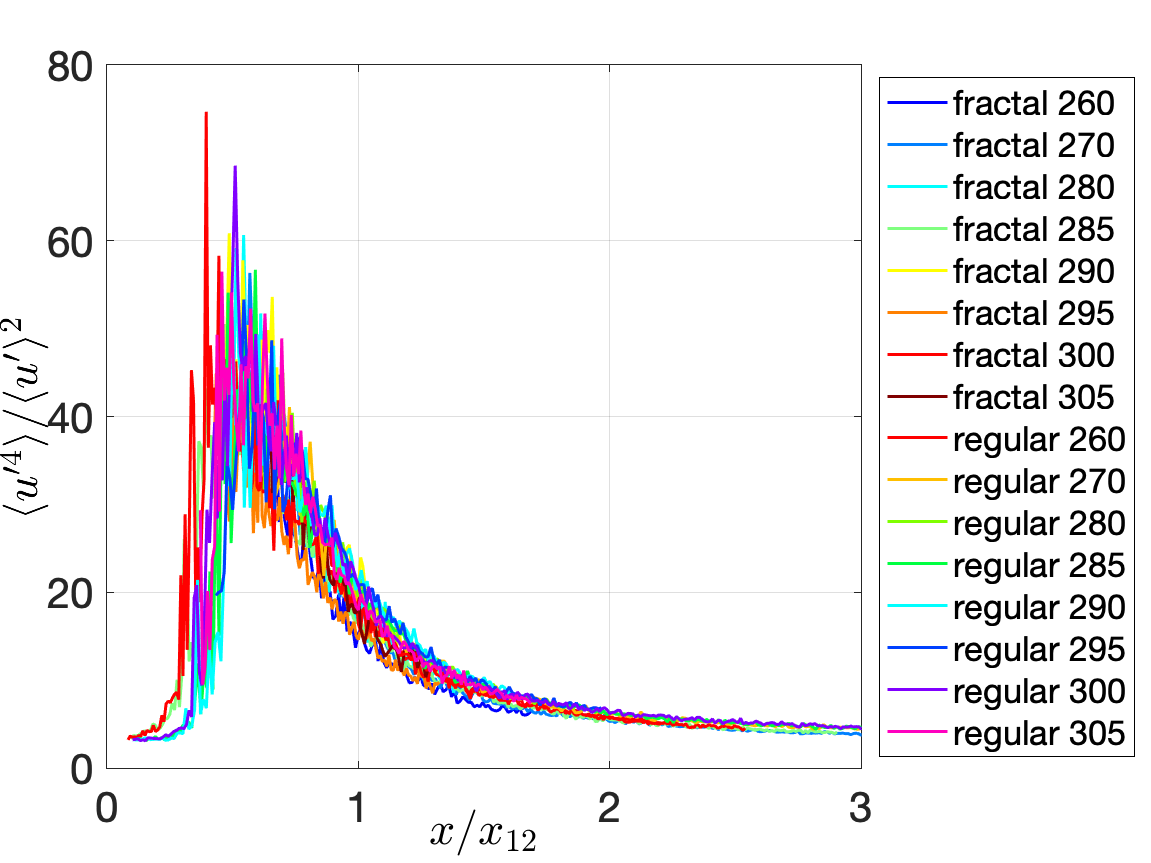}}
\caption{Scaled skewness and flatness of irregular and regular plates}
\end{figure}

\begin{figure}
\centering
\includegraphics[width=0.9\textwidth]{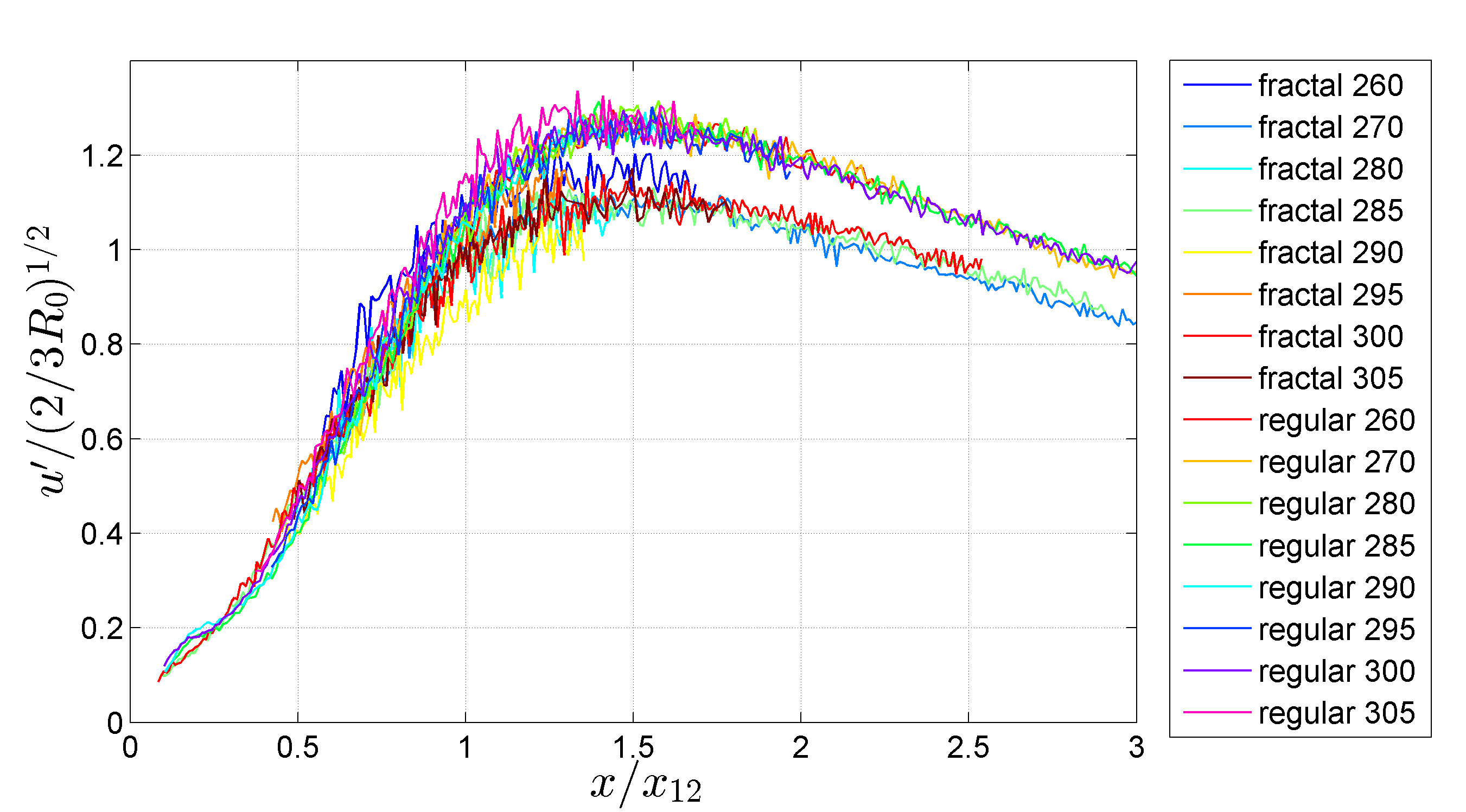}
\caption{Scaled $u\prime$ of the irregular and regular square plates.}
\label{abb_31_All_Uprime_collapsed}
\end{figure}

The major results from the irregular and the regular square plates are
now compared. We focus on $S\geq260$mm, where the profile collapse is
best in terms of $x_{12}$. In figure \ref{abb_24_ALL_SKEW} and
\ref{abb_24_ALL_FLAT} we plot the streamwise profiles of the skewness
and the flatness of the turbulent velocity fluctuations scaled by the
interaction length scale $x_{12}$ for both the irregular and the
regular set of square plates. A reasonable collapse is achieved. In
figure \ref{abb_31_All_Uprime_collapsed}, the streamwise profiles of
the scaled velocity fluctuations are displayed. It can be seen that
the maxima of $u^\prime / \sqrt{\left( 2/3 R_0 \right)}$ (where $R_0 =
U_\infty u_0(d/dx\,\delta(x))$) are higher for the regular plates than
for the irregular ones. The scaling $u\prime \sim \sqrt{U_\infty
  u_0(d/dx\,\delta(x))} F(x/x_{12})$ seems to hold for both sets of
plates, but the proportionality constant is different depending on
plate geometry.
%

\section{Conclusions}

In this work we have studied experimentally via hot-wire anemometry
the interaction between two turbulent axisymmetric wakes in the wind
tunnel. The wake generators were pairs of plates, one pair with plates
of regular peripheries and one pair with plates of irregular
peripheries. We first pointed out that the wake interaction length
defined from $2\delta (x^{*}) \propto S$ using the wake half width
$\delta$ and the separation $S$ between the two plates depends on the
nature of the turbulent kinetic energy cascade. For the case of the
standard Richardson-Kolmogorov energy cascade, $x^*$ evolves as $S^3$
while a non-equilibrium cascade implies $x^* \propto S^2$.

By acquiring streamwise profiles for different plate separations and
identifying the wake interaction length for each separation it is
possible to show that the interaction between the wakes is consistent
with non-equilibrium scalings. The profile of the streamwise
distribution of the normalised velocity fluctuations $u\prime/\langle
U(x)\rangle$ is used to characterise the interaction of two wakes, as
it shows three clearly defined regions: a first region where the wakes
have not significantly met yet and the flow is highly intermittent and
a second region where the wakes start to merge and where
$u\prime/\langle U(x)\rangle$ reaches a maximum value. Finally, in the
third region both wakes are fully merged and the fluctuations exhibit
a monotonic decrease with streamwise distance.

We have therefore proposed to identify the wake interaction length
with the streamwise point $x_{12}$ where the flow goes from region 1
to region 2. We found that the values of $x_{12}$ are indeed
consistent with the non-equilibrium cascade, as $x_{12} \propto
S^2$. We also defined a wake interaction length-scale independently
from $x_{12}$ and have shown that it is proportional to $x_{12}$.  We
then used $x_{12}$ to successfully collapse the streamwise profiles of
the second, third and fourth moments of the streamwise fluctuating
velocity.

Following previous theoretical developments
(\cite{townsend80,george89, dairay15}) which demonstrated that the
turbulent kinetic energy and the Reynolds shear stress both scale with
streamwise distance as $U_\infty u_0(d/dx\,\delta(x))$ we have
proposed the scaling $u\prime \sim \sqrt{U_\infty
  u_0(d/dx\,\delta(x))} F(x/x_{12})$ where $F$ is a dimensionless
function of $x/x_{12}$. This scaling holds for both sets of plates,
but the proportionality constant is larger for the regular than the
irregular plates.

Our results and analysis could contribute to the study of interactions
between wakes of neighboring wind turbines and the design of wind
farms.

\subsection*{Acknowledgements}
We acknowledge the support of ERC Advanced Grant 320560 awarded to JCV. The authors report no conflict of interest.

\bibliographystyle{jfm}

\end{document}